\documentclass[aps,prx,reprint,superscriptaddress,showpacs]{revtex4-1}
\usepackage{color}
\usepackage{graphicx}
\usepackage[american]{babel}
\usepackage[T1]{fontenc}
\usepackage{amsmath}
\usepackage{amssymb}
\usepackage{amstext}
\usepackage{amsthm}
\usepackage{latexsym}
\usepackage{verbatim}
\usepackage{natbib}
\usepackage{array}
\usepackage{color} 

\usepackage{ulem}   

\def\e{\varepsilon}
\def\d{\delta}

\def\l{\lambda}

\def\t{\tau}

\renewcommand{\o}{\omega}

\def\s{\sigma}

\def\G{\Gamma}

\renewcommand{\O}{\Omega}
\def\L{\Lambda}

\def\ra{\rightarrow}
\def\up{\uparrow}

\def\down{\downarrow}

\def\Ra{\Rightarrow}

\def\bk{{\bf k}}

\def\bq{{\bf q}}

\def\bQ{{\bf Q}}

\def\bS{{\bf S}}

\newcommand{\be}{\begin{equation}}
\newcommand{\ee}{\end{equation}}
\newcommand{\bea}{\begin{eqnarray}}
\newcommand{\eea}{\end{eqnarray}}
\def\pref#1{(\ref{#1})}
\newcount\bozza \bozza=0
\ifnum\bozza=1
\newdimen\shift \shift=-2truecm
\def\lb#1{%
{\label{#1}\rlap{\kern\shift{$\scriptstyle#1$}}}}
\else\def\lb#1{\label{#1}} \fi

\newcommand{\nn}{\nonumber}

\begin{document}

\title{Orbital-dependent Fermi Surface shrinking as a fingerprint of nematicity in FeSe}

\author{Laura Fanfarillo*}
\affiliation{CNR-IOM and International School for Advanced Studies (SISSA),
Via Bonomea 265, I-34136, Trieste, Italy}
\author{Joseph Mansart*}
\affiliation{Laboratoire de Physique des Solides, Universit\'{e} Paris-Sud, UMR 8502, B\^at. 510, 91405 Orsay, France}

\author{Pierre Toulemonde}

\author{Herv\'e Cercellier}
\affiliation{Univ. Grenoble Alpes, Institut NEEL, F-38000 Grenoble, France}
\affiliation{CNRS, Inst. NEEL, F-38000 Grenoble, France}

\author{Patrick Le F\`evre}

\author{Fran\c{c}ois Bertran}
\affiliation{Synchrotron SOLEIL, L'Orme des Merisiers, Saint-Aubin-BP 48, 91192 Gif sur Yvette, France}

\author{Belen Valenzuela}
\affiliation{Instituto de Ciencia de Materiales de Madrid, ICMM-CSIC, Cantoblanco, E-28049 Madrid, Spain}

\author{Lara Benfatto}
\affiliation{ISC-CNR and Dep. of Physics, ``Sapienza'' University of Rome, P.le A. Moro 5, 00185, Rome, Italy}

\author{V\'eronique Brouet $^{\dagger,} $}
\affiliation{ Laboratoire de Physique des Solides, Universit\'{e} Paris-Sud, UMR 8502, B\^at. 510, 91405 Orsay, France}

\begin{abstract}
\noindent * These authors contributed equally to the work\\
$\dagger$ corresponding author : veronique.brouet@u-psud.fr\\

The large anisotropy in the electronic properties across a structural transition in several correlated systems has been identified as the key manifestation of electronic nematic order, breaking rotational symmetry. In this context, FeSe is attracting tremendous interest, 
since electronic nematicity develops over a wide range of temperatures, allowing accurate experimental investigation. Here we combine 
angle-resolved photoemission spectroscopy and theoretical calculations based on a realistic multi-orbital  model to unveil the microscopic mechanism responsible for the 
evolution of the electronic structure of FeSe across the nematic transition. We show that the  self-energy corrections due to the exchange of spin fluctuations between hole and electron pockets are responsible for an orbital-dependent shrinking of the Fermi Surface that affects mainly the $xz/yz$ parts of the Fermi surface. This result is consistent with our experimental observation of the Fermi Surface in the high-temperature tetragonal phase, that  includes the $xy$ electron sheet that was not clearly resolved before. In the low-temperature nematic phase, we experimentally confirm the appearance of a large ($\sim$ 50meV) $xz/yz$ splittings. It can be well reproduced in our model by assuming a moderate splitting between spin fluctuations along the $x$ and $y$ crystallographic directions. Our mechanism shows how the
full entanglement between orbital and spin degrees of freedom can make a spin-driven nematic transition equivalent to an effective orbital order. 
\end{abstract}

\date{\today}

\maketitle

\section{Introduction}
Electronic nematic phases are ordered states, where electrons spontaneously break the rotational point-group symmetry of the crystal, but not its translational symmetry. They are increasingly believed to play an important role in many correlated systems \cite{FradkinReview2010}. In recent years, iron based superconductors have provided remarkable examples for such behaviors \cite{FisherRPP2011,FernandesNatPhys14,GallaisReview15}. Indeed, the electronic properties manifest a much larger anisotropy across the tetragonal to orthorhombic transition than expected from the structural changes alone. One possibility is that the nematic phase is a precursor of the antiferromagnetic order that usually emerges at lower temperature by selecting an ordering wave vector along the $x$ direction. However, direct measurements of the band structure seem to point to a true symmetry-breaking state with a charge unbalance between the $xz$ and $yz$ orbitals. As these degrees of freedom are strongly entangled, it is not easy to discriminate their respective role.

FeSe offers the unique opportunity to study the nematic behavior occurring below the structural transition at T$_S$$\sim$90K in the absence of any long-range magnetic ordering \cite{BohmerPRB13}. A superconducting state eventually develops below 9K in bulk FeSe samples \cite{HsuPNAS08}. In this wide temperature range, the system shows a marked electron nematicity in transport \cite{Tanatarcm15}. Angle resolved photoemission spectroscopy (ARPES) investigations \cite{ShimojimaPRB14,NakayamaPRL14,WatsonPRB15,ZhangDingPRB15,ZhangShen15} have revealed a 50meV splitting at the M point of the Brillouin Zone (BZ) between $xz$ and $yz$ orbitals. However, this splitting is different at $\Gamma$ \cite{ZhangDingPRB15,ZhangShen15} and actually of {\it opposite} sign \cite{Suzuki15}. This rules out a simple on-site unbalance of orbital occupation \cite{LvPRB09} (ferro-orbital order) and suggests instead the emergence of momentum-modulated orbital ordering \cite{SuJcondMat15,MukherjeePRL15,JiangCondMat15}, whose microscopic origin still remains debated. Even though the spin-driven nematic scenario was first considered as unlikely due to the lack of long-range magnetic order or precursor effects\cite{BaekNatMat14,BohmerPRL15}, more recently sizable spin fluctuations have been detected\cite{Wang15,Rahn15,Shamoto15,VasilievCondMat16}. In addition, it has been pointed out that the absence of long-range magnetic order could be due to frustration\cite{Glasbrenner15}, leaving open the possibility that fluctuating magnetism can play a role also in FeSe, as it occurs in other iron-based systems. 

In addition to the above findings, FeSe exhibits, already at high temperatures well above $T_S$, a dramatic \lq\lq{}shrinking\rq\rq{} of the FS pockets as compared to local density approximation (LDA) calculations. This means that bands are shifted in {\it opposite} downward/upward directions for hole pockets at $\Gamma$ and electron pockets at M, respectively. This has been observed previously in a number of iron pnictides, by quantum oscillations \cite{coldeaPRL08} or ARPES \cite{BrouetPRL13}, but is rarely considered as an important fingerprint of interactions in these systems. Dynamical mean-field theory (DMFT) calculations for example reproduce remarkably well the observed mass renormalizations in these systems \cite{Demedici14,AichhornPRBFeSe}, but they do not predict a strong shrinking in general, and not in FeSe \cite{AichhornPRBFeSe,YinPRB12}. On the other hand, the exchange of spin fluctuations between hole and electron pockets can provide a general mechanism for the FS shrinking in pnictides \cite{OrtenziPRL09,BenfattoPRB11}. In this paper, we extend this previous approach \cite{OrtenziPRL09} to a realistic microscopic model for spin interactions where the spin-fluctuation exchange mechanism is orbital-selective, as recently pointed out in Ref.\ \cite{FanfarilloPRB15}. We then argue that an orbital-dependent shrinking of the Fermi surfaces is the key mechanism to understand the nematic transition in FeSe. Experimentally, we achieve a clear identification in ARPES measurements of the $xy$ electron band. This allows us to establish that above $T_S$ the FS shrinking is stronger than in other pnictides and it is orbital-selective, the $xz/yz$ FS sheets being much more severely affected than the $xy$ sheets.  Below $T_S$, the $xy$ electron band is basically unaffected, supporting the original\cite{ShimojimaPRB14,NakayamaPRL14,WatsonPRB15,ZhangDingPRB15,ZhangShen15} view - questioned in some recent reports\cite{Watson2016,BorCondMat16}-  that the band-structure modifications arise from a 50meV splitting of the $xz-yz$ orbitals. This energy splitting of the $xz/yz$ orbitals directly follows from an orbital differentiation of the FS shrinking mechanism. At microscopic level this is due to the anisotropy of the spin fluctuations peaked at ordering vectors along $k_x$ or $k_y$, a fingerprint of the spin-driven nematic scenario \cite{FernandesNatPhys14}. Our picture not only creates a strong link between the FS shrinking and nematicity, able to describe the ARPES data above and below the structural transition, but it also solves the apparent dichotomy between spin-driven and orbital-driven nematic scenarios, that merge in our approach in an unified orbital-selective spin-fluctuation nematic mechanism. 


\section{Experimental details}

Single crystals have been grown using the chemical vapor transport method in sealed quartz tube, starting from Fe and Se powders (with a 1.1 : 1 molar ratio) in an eutectic KCl+AlCl$_3$ chlorides mixture. Details and characterization can be found in ref. \cite{Karlsson15}. The observation of quantum oscillations \cite{AudouardEPL14} attests from the samples\rq{}quality.  

ARPES measurements were carried out at the CASSIOPEE beamline of the SOLEIL synchrotron, with a Scienta R4000 analyzer, an angular resolution of 0.3$^{\circ}$ and an overall energy resolution better than 10 meV. The measurements in Fig. 2 and 4 were carried out at a photon energy of 40eV with linear polarization along $k_x$. This selects even orbitals along $k_x$ and odd orbitals along $k_y$. More details are given in SI. 

\section{Results and discussion}

\subsection{Model of orbital-selective shrinking.} 

\begin{figure}[tbp]
 \centering
 \includegraphics[width=0.48\textwidth]{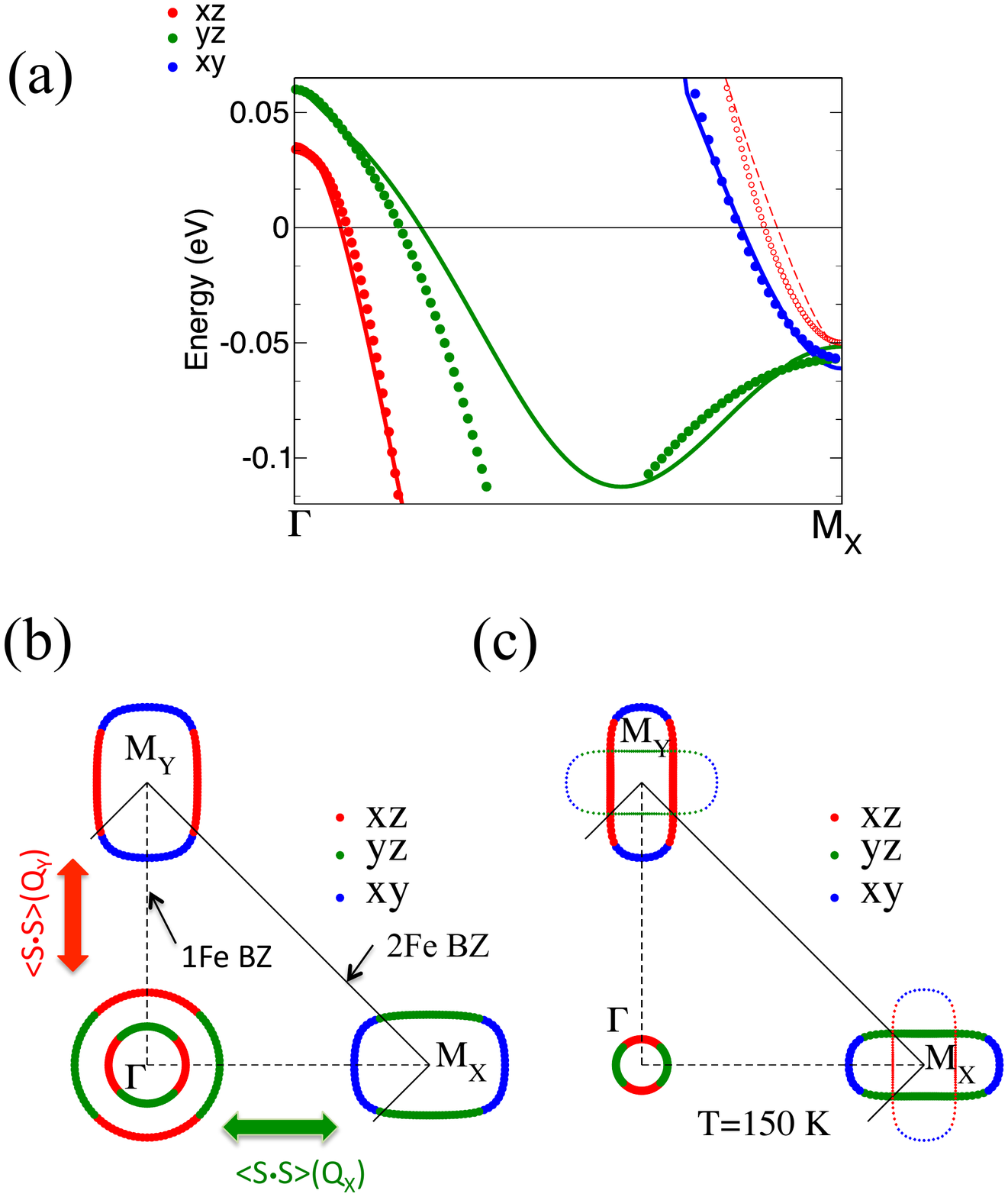}
 \caption{General sketch of the electronic structure. (a) Lines : Band structure of FeSe at $k_z=0$ along the $k_x$ direction of the 1 Fe BZ obtained with a tight-binding model including SOC and mass renormalization (see text). Symbols show the low-energy model that we use in the calculation [Eq.\ \pref{h0gen}]. Dashed lines and open symbols denote the bands folded at $\G$ and $M_X$ in the 2Fe BZ. (b) FS cuts for the low-energy model in the 1Fe BZ. The exchange of spin flucctuations between hole and electron pockets at $M_X$ or $M_Y$ shrinks selectively the $yz$ or $xz$ orbitals, respectively.  This effect, that is already present above $T_S$, modifies significantly the FS, as shown in panel (c) at 150 K. Here small symbols denote the bands folded in the 2Fe BZ.}
 \label{Fig1_model}
 \end{figure}

As a starting point, we model the band structure with the tight-binding model of Ref. \cite{EshrigPRB09} (Fig.\ \ref{Fig1_model}a), where the renormalization of the bands due to Hubbard and Hund's like interactions is already taken into account. As it will be justified by our ARPES data later on, this requires a high-temperature renormalization of 3 for $xz/yz$ and 5 for $xy$ orbitals, in good agreement with DMFT calculations \cite{AichhornPRBFeSe,Demedici14}. To allow for an analytical treatment we map this dispersion into a 
low-energy two-dimensional model able to describe the relevant orbital content of the pockets around the $\G$, $M_X$ ($\pi$,0) and $M_Y$ (0,$\pi$) points in the 1Fe BZ \cite{VafekPRB13}. The Hamiltonian at each point can be represented as $H_0^l=\sum_{\bk,\s} \Psi^{l,\dagger}_{\bk\s} \hat H_0^l \Psi^l_{\bk\s}$ where $\hat H_0^l$ ($l=\G,M_X,M_Y$) is a $2\times2$ matrix and the spinors are defined as $\Psi^\G_{\bk\s}=(c^{yz}_{\bk\s},c^{xz}_{\bk\s})$ and $\Psi^{X/Y}_{\bk\s}=(c^{yz/xz}_{\bk\s},c^{xy}_{\bk\s})$. The additional $xy$ hole pocket at $\Gamma$ is not included since it is below $E_F$, as confirmed by previous ARPES measurements\cite{WatsonPRB15,ZhangDingPRB15,BorCondMat16} and by our data at low temperature, see below. The matrix $\hat H^l$ has the general structure 
\be
\lb{h0gen}
\hat H_0^{l}=
h_0^l\t_0+\vec{h}^l\cdot\vec{\t}^l
\ee
where $\vec{\t}$ are Pauli matrices representing the orbital isospin. The bands $E^{l\pm}_\bk=h_0^l\pm|\vec h^l|$ and their orbital content are then simply deduced by a straightforward diagonalization of the Hamiltonian \pref{h0gen}.  By using the explicit expressions of $(h_0,\vec{h})$ detailed in the Supplementary Information (SI) one obtains the approximate band dispersions shown by symbols in Fig.\ \ref{Fig1_model}a and giving the FS shown in Fig.\ \ref{Fig1_model}b. At the $\G$ point we added 
explicitly to the Hamiltonian \pref{h0gen} the spin-orbit coupling (SOC) $\lambda$\cite{VafekPRB13,FernandesSO} via the replacement of $|\vec h^\G(\bk)|$ with $\sqrt{|\vec h^\G(\bk)|^2+\l^2/4}$. This lifts the degeneracy of the inner and outer $xz/yz$ pockets at $\Gamma$. 

The basic mechanism of the FS shrinking developed in Ref.\ \cite{OrtenziPRL09,BenfattoPRB11} focuses on the changes of the low-energy effective model induced by the coupling to collective modes, described within an Eliashberg framework via a self-energy function $\Sigma^l(\o)$ for each band. 
The strong particle-hole asymmetry of the bands in pnictides leads to a {\it finite} real part $\zeta^l(\o)$ of the self-energy $\Sigma^l(\o)$, responsible for an energy-dependent shift of the interacting bands. In particular its sign in the pocket $l$ is determined by 
$\zeta^l\simeq -\ln|E^{l'}_{top}/E^{l'}_{bot}|$, where $E^{l'}_{top}$, $E^{l'}_{bot}$ are the energy difference between the top/bottom of the $l'$ pocket from the Fermi level \cite{OrtenziPRL09}.  Thus, when the exchange mechanism is {\it interband}, the sign of $\zeta^\G(\o)$ is controlled by the electron pockets at $M$, having $E_{top}\gg E_{bot}$, and it is thus negative. Conversely, for the hole bands $E_{top}\ll E_{bot}$ and the induced shift on the electron pockets is positive. In both cases one finds a {\it shrinking} of the FS, in agreement with observations in several iron-based pnictides \cite{coldeaPRL08,BrouetPRL13}. The most natural bosonic mode responsible for this effect are then spin fluctuations (SF) $\langle \bS\cdot \bS\rangle (\bQ)$ at momenta $\bQ_X\equiv \G M_X$ or $\bQ_Y\equiv \G M_Y$ connecting hole bands at $\Gamma$ with electron bands at $M_X, M_Y$, see Fig.\ \ref{Fig1_model}b. On this respect, our self-energy shrinking mechanism stems somehow as a low-energy counterpart of the so-called $s^\pm$ Pomeranchuk instability that has been found, by renormalization-group approaches\cite{Lee_pom_prb09,ChubukovCondMat16}, as a possible competing instability triggered by the proximity to a spin-density wave order. Indeed, while the particle-hole asymmetry only guarantees a finite value of $\zeta^l(\o)$ when the carriers are coupled to a bosonic mode, is the interband nature of the mode, i.e. its identification with spin fluctuations, that guarantees a band shift reducing the FS areas.

 \begin{figure*}[tbp]
 \centering
 \includegraphics[width=0.9\textwidth]{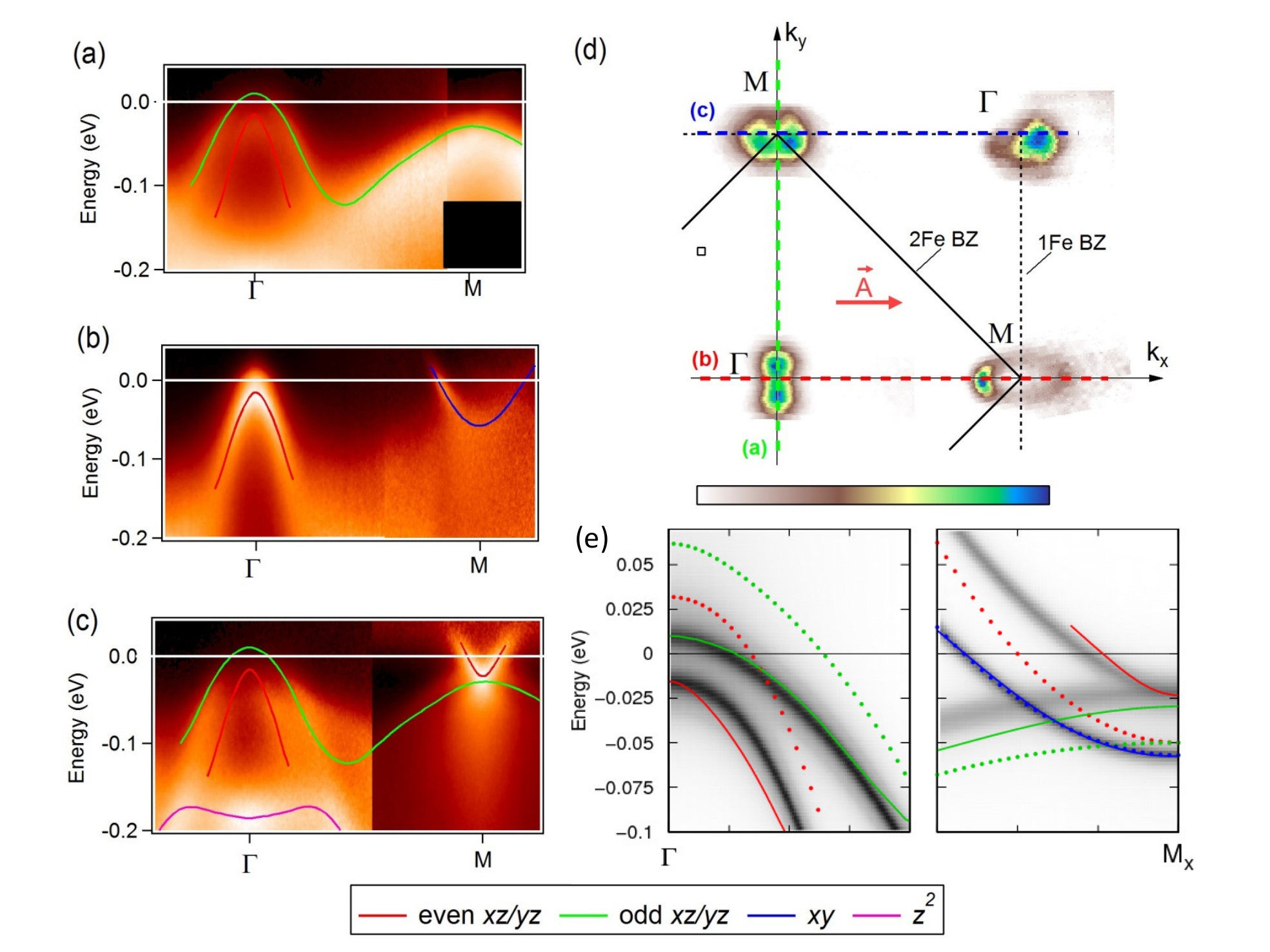}
 \caption{Electronic structure at 150K measured by ARPES. (a-c) Energy 
 momentum plots along three $\Gamma$M directions indicated as thick dashed lines 
 in panel (d). Lines are guides to the eyes indicating the dispersions of the 
 different bands, with colors encoding the main orbital character (the even 
 $xz/yz$ is $xz$ along $k_x$ and $yz$ along $k_y$). The data were measured at 
 40eV photon energy ($k_z$$\sim$0) with linear polarization along $k_x$. In (c), 
 the area at each energy have been normalized to enhance the visibility of the 
 electron pocket. (d) Fermi Surface map obtained by integration of the ARPES 
 spectral weight at +10meV in a 4meV window. (e) Image plot in gray scale of the 
 spectral functions of the renormalized bands at 150 K obtained including 
 self-energy corrections. Ticks along abscissa corresponds to 0.1 $\Gamma$M. Thin lines follow the experimental data shown in (a-c). As in Fig \ref{Fig1_model}a, symbols indicate the bare bands of Eq.\ \pref{h0gen}.} \label{Fig_150K}
 \end{figure*}

So far, the FS shrinking mechanism has not been linked to the orbital degrees of freedom. To this extent, two additional ingredients should be added to the previous approach: (i) computing explicitly the self-energy effect within the orbital model  \pref{h0gen}, instead of the band model considered in Ref.\ \cite{OrtenziPRL09}, and (ii) accounting for the fact, pointed out recently in Ref.\ \cite{FanfarilloPRB15}, that the mechanism of SF exchange must preserve the orbital character of the electrons. Thanks to the item (ii) one can show  (see SI) that the self-energy matrix  $\hat \Sigma^l$, relating via the Dyson equation $(\hat G^l)^{-1}=(\hat G^l_0)^{-1}-\hat \Sigma^l$  the bare $\hat G^l_0$  and the dressed $\hat G^l$ matrix  Green's functions of the model \pref{h0gen}, simplifies considerably :
\be
\lb{smatrix}
\hat \Sigma^\G=\begin{pmatrix}
\Sigma^\G_{yz} & 0\\
0 &\Sigma^\G_{xz} \\
\end{pmatrix},
\hat \Sigma^{X/Y}=\begin{pmatrix}
\Sigma^{X/Y}_{yz/xz} & 0\\
0 &0 \\
\end{pmatrix}
\ee
where $\Sigma^{X,Y}_{xy}=0$ is a consequence of the lack of an $xy$ orbital component on the hole pockets. More importantly, we will show below that the self-energy functions select SF around precise wave-vectors, linking the $yz$ orbital with SF at $Q_X$ and $xz$ with SF at $Q_y$:
%
\bea
\lb{corrx}
\Sigma^{\G}_{yz}(\o), \Sigma^{X}_{yz}(\o) &\Ra& \langle \bS\cdot \bS\rangle (\bQ_X), \\
\lb{corry}
\Sigma^{\G}_{xz}(\o),  \Sigma^{Y}_{xz}(\o) &\Ra& \langle \bS\cdot \bS\rangle (\bQ_Y)
\eea
This result basically follows from the fact that in the 1Fe BZ,  {\it only $yz$ is present in the electron pocket at $\bQ_X$, and only $xz$ is present at $\bQ_Y$.} Here the basic mechanism \cite{OrtenziPRL09} controlling the sign of the self-energy corrections remains unchanged, since SF always connect hole and electron pocket. In addition, the result
\pref{corrx}-\pref{corry}  translates the inequivalence \cite{FernandesNatPhys14} of SF at $\bQ_X$ and $\bQ_Y$ below $T_S$ into inequivalent corrections for $yz$ and $xz$ orbitals. Therefore, even though no electronic order parameter develops at $T_s$, neither in the charge nor in the spin sector, the anisotropy of the spin fluctuations, that is the hallmark of a spin-nematic transition, induces an anisotropy of the self-energy corrections that acts as an effective orbital ordering. Eq.s\ \pref{smatrix}-\pref{corry} contain the essence of  the orbital-selective shrinking that we will discuss in the following. 


\subsection{Isotropic Fermi Surface shrinking at 150K.} 

In Fig. \ref{Fig_150K}, we 
present the electronic structure of FeSe measured by ARPES at 150K, well above 
the structural transition. As detailed in SI, to observe all the bands by ARPES, 
it is necessary to combine even/odd light polarizations and measurements in 
different Brillouin zones (BZ). Three different versions of $\Gamma$M are 
presented in panels (a-c) to cover all bands, corresponding to the three cuts 
indicated on the FS map (d) by thick dashed lines. The different dispersions 
are modeled by thin lines which are guides to the eyes, and are also reported in 
Fig.\ref{Fig_150K}(e). The colors indicate the main orbital character (see 
caption).  

The general structure of the bands is in agreement with the one outlined in Fig.\ \ref{Fig1_model}. Around the $\Gamma$ point, we observe two hole-like bands, made by $xz$ and $yz$ orbitals. They are splitted at $\Gamma$ by 20 meV, which we attribute to SOC \cite{MaletzPRB14,WatsonPRB15}. 
The odd $xz/yz$ orbital (i.e. $yz$ along $k_x$ and $xz$ along $k_y$) forms a \lq\lq{}saddle\rq\rq{} band at the M point, where it is expected to be degenerate with a shallow electron pocket of opposite $xz/yz$ character. Our best fit gives a small residual splitting of $\sim$5~meV at M, although it could be within error bars. Note that the SOC is not effective at M between these $xz$ and $yz$ bands, because they are formed by different combinations of the 2 Fe of the unit cell (see Fig. 1(b) \cite{HaoPRX14,FernandesSO}).  

\begin{figure*}[tbp]
 \centering
 \includegraphics[width=0.9\textwidth]{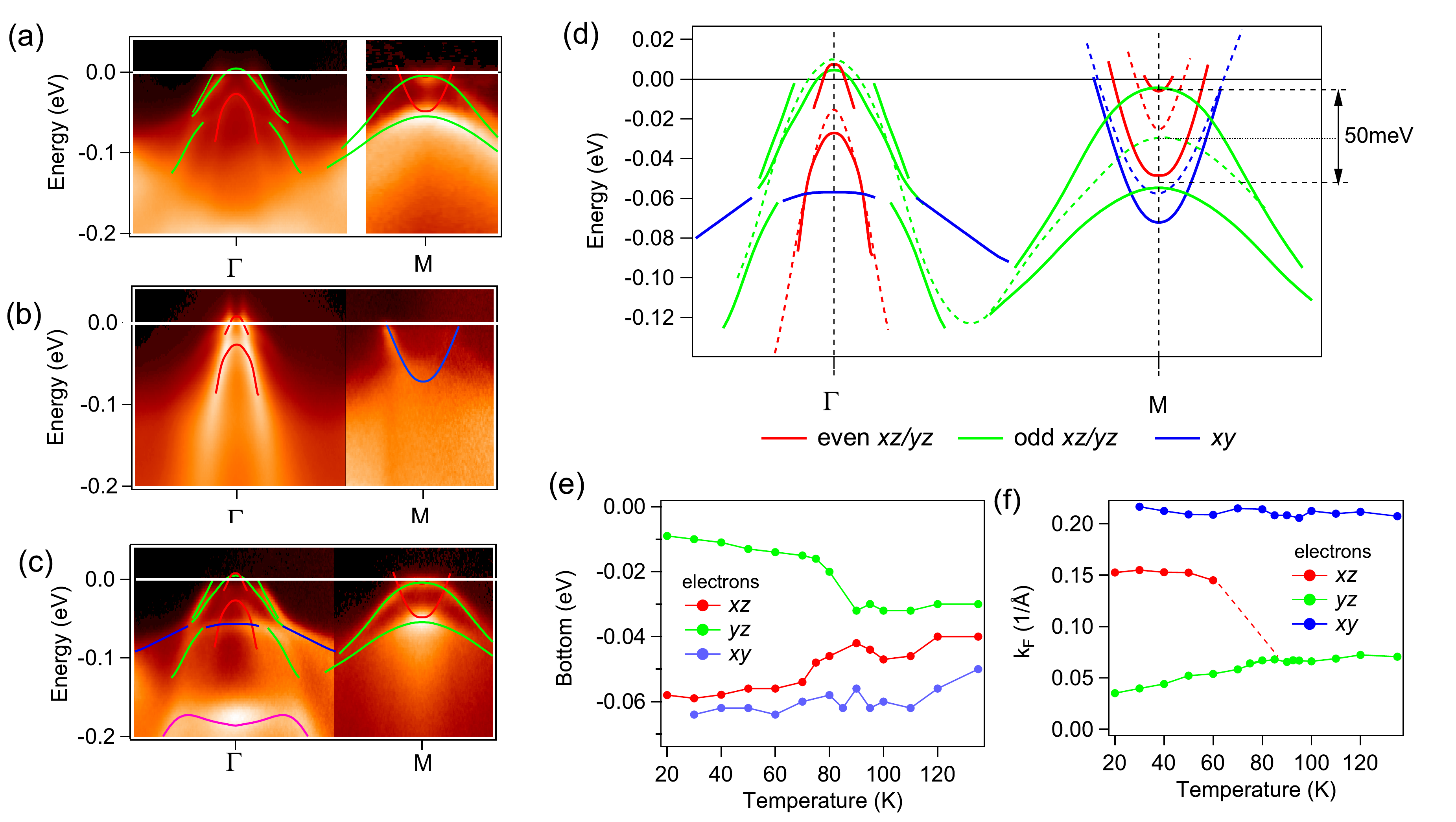}
 \caption{Electronic structure at 20K measured by ARPES. (a-c) Energy 
 momentum plots in the same conditions as Fig.\ref{Fig_150K}, but at 20K. (d) Thin color line: experimental guides of the dispersions, also indicated in (a-c). Dashed lines : experimental guides at 150K. (e-f) 
 Evolution with temperature of the bottom (e) and $k_F$ (f) of the three electron 
 bands. } \label{Fig_20K}
 \end{figure*}

Another, deeper, electron band around M, appearing only in Fig. \ref{Fig_150K}(b), has $xy$ character. It is often difficult to observe it in iron pnictides and was not reported before for FeSe at high temperatures. As we mentioned above, the hole-like counterpart of $xy$ at $\Gamma$ is not very visible at this temperature, although it will be clearer at lower temperatures and 60meV below $E_F$ (see Fig. \ref{Fig_20K}). 

In Fig. \ref{Fig_150K}(e), we see that the slope of the experimental dispersions (thin lines) compare well with the renormalization values assumed in Fig.\ \ref{Fig1_model}a (symbols). However, the Fermi wave-vectors $k_F$ corresponding to the $xz/yz$ orbitals are clearly too large, both for hole and electron pockets. On the contrary, it is approximately correct for the $xy$ electron band. To fully appreciate the amplitude of the shrinking, we estimate in SI the FS volume after integration over $k_z$. We find that it is reduced by a factor 5 to 10 for the $xz/yz$ parts. This is the largest shrinking observed in iron based superconductors to our knowledge. We estimated a factor 2 in Co-doped BaFe$_2$As$_2$ \cite{BrouetPRL13} and at most 1.2 in LiFeAs \cite{BrouetLiFeAs}. Notice that in LiFeAs there is a strong shrinking of the hole $xz/yz$ pockets, but it is compensated by an expansion of the $xy$ hole band \cite{BrouetLiFeAs,LeePRL12}, so that the total number of carriers and the size of the electron pockets is nearly unchanged \cite{BrouetLiFeAs}. This case is however completely different from the shrinking considered here, involving a mutual compensation between hole and electron pockets. In addition, while the orbital redistribution of holes between the $xz/yz$ and $xy$ sheets in LiFeAs is well captured by DMFT\cite{LeePRL12},  the same effect has not been reported by DMFT calculations in FeSe \cite{AichhornPRBFeSe,YinPRB12}. 

As shown in Fig.\ \ref{Fig_150K}e this orbital-selective shift of the bands is a natural outcome of the self-energy effects encoded in Eq.\ \pref{smatrix}. As discussed previously\cite{OrtenziPRL09,BenfattoPRB11}, to capture the basic ingredients of the FS shrinking we can discard the full momentum dependence of the SF propagator, and use the form
\be
\lb{bapp}
B_{X/Y}(\o)=\frac{1}{\pi}\frac{\o\o_{0}}{(\o^{X/Y}_{sf}(T))^2+\O^2}, 
\ee
where $\omega_0$ is a constant while $\o^{X/Y}_{sf}(T)$   is the characteristic  energy scale of spin modes. In the tetragonal phase we assumed the typical temperature evolution of the paramagnetic SF, $\o^{sf}_{X,Y}(T)=\o_0(1+T/T_{\theta})$, as observed above $T_c$ in pnictide systems\cite{InosovNatPhys10}. Here we used $\o_0\sim 20$ meV and $T_\theta\sim150$ K, in agreement with experimental results in 122 systems and more recently also in FeSe\cite{Rahn15}.
The self-energy functions appearing in Eq.s \pref{smatrix} are then computed as:
\bea
\lb{sigmagx}
\Sigma^{\G}_{yz}(i\o_n)&=& - V T\sum_{\bk,m}  D_X(\o_n-\o_m)g^X_+(\bk,i\o_m)\\
\lb{sigmagy}
\Sigma^{\G}_{xz}(i\o_n)&=& - V T\sum_{\bk,m}  D_Y(\o_n-\o_m)g^Y_+(\bk,i\o_m)
\eea
where $D_{X/Y}(\omega_n)= \int d\Omega~2\Omega B_{X/Y}(\Omega)/(\Omega^2+\omega_n^2)$ is the propagator for SF along $k_x/k_y$,  $B_{X/Y}$ is its spectral function given by Eq.\ (\ref{bapp}) above, $V$ is the strength of the coupling and $g^l_\pm(\bk,i\o_m)$ denotes the Greens function of the $E^{l,\pm}$ band at the $l$ pocket (more details are given in SI).  Analogously for the $X,Y$ pockets one has
\bea
\Sigma^{X}_{yz}(i\o_n)&=& - V T\sum_{\bk,m}  D_{X}(\o_n-\o_m)(g^\G_+(\bk,i\o_m)+g^\G_-(\bk,i\o_m))\nn\\
\lb{sigmax}\\
\Sigma^{Y}_{xz}(i\o_n) &=& - V T\sum_{\bk,m}  D_{Y}(\o_n-\o_m)(g^\G_+(\bk,i\o_m)+g^\G_-(\bk,i\o_m)).\nn\\
\lb{sigmay}
\eea

Above $T_S$ SF are isotropic in momentum space, i.e. $\omega_{sf}^X=\omega_{sf}^Y$,  so that the self-energy corrections \pref{corrx}-\pref{corry} are isotropic in the orbital space, but they have opposite signs on the hole and electron pockets (see also Fig.\ \ref{Fig_SF}(c-d) below). Indeed, as discussed above, the real parts of $\Sigma^{\G}_{yz/xz}$ are negative, as due to the electron-like particle-hole asymmetry of the electron pockets appearing on the r.h.s. of Eq.\ \pref{sigmagx}-\pref{sigmagy}, and conversely those of $\Sigma^{X/Y}_{yz/xz}$ are positive. To further account for the  different degrees of nesting of the various pockets we modulated the couplings $V$ in the above equations, as detailed in SI. Because $\Sigma^{X/Y}_{xy}=0$, as stated above, there is no shrinking on $xy$. As a consequence, while at $\G$ the pockets change their size but not their shape, the electron bands become more elliptical (see Fig \ref{Fig1_model}c) in agreement with the experiments.  We stress that because of the frequency dependence of the self-energy, the present mechanism of FS shrinking is not equivalent to a rigid band shift \cite{BenfattoPRB11}, even though this cannot be easily appreciated on the energy scale of Fig.\ \ref{Fig_150K}e.

\subsection{Anisotropic Fermi Surface shrinking at 20K.}

 \begin{figure*}[tbp]
 \centering
 \includegraphics[width=0.9\textwidth]{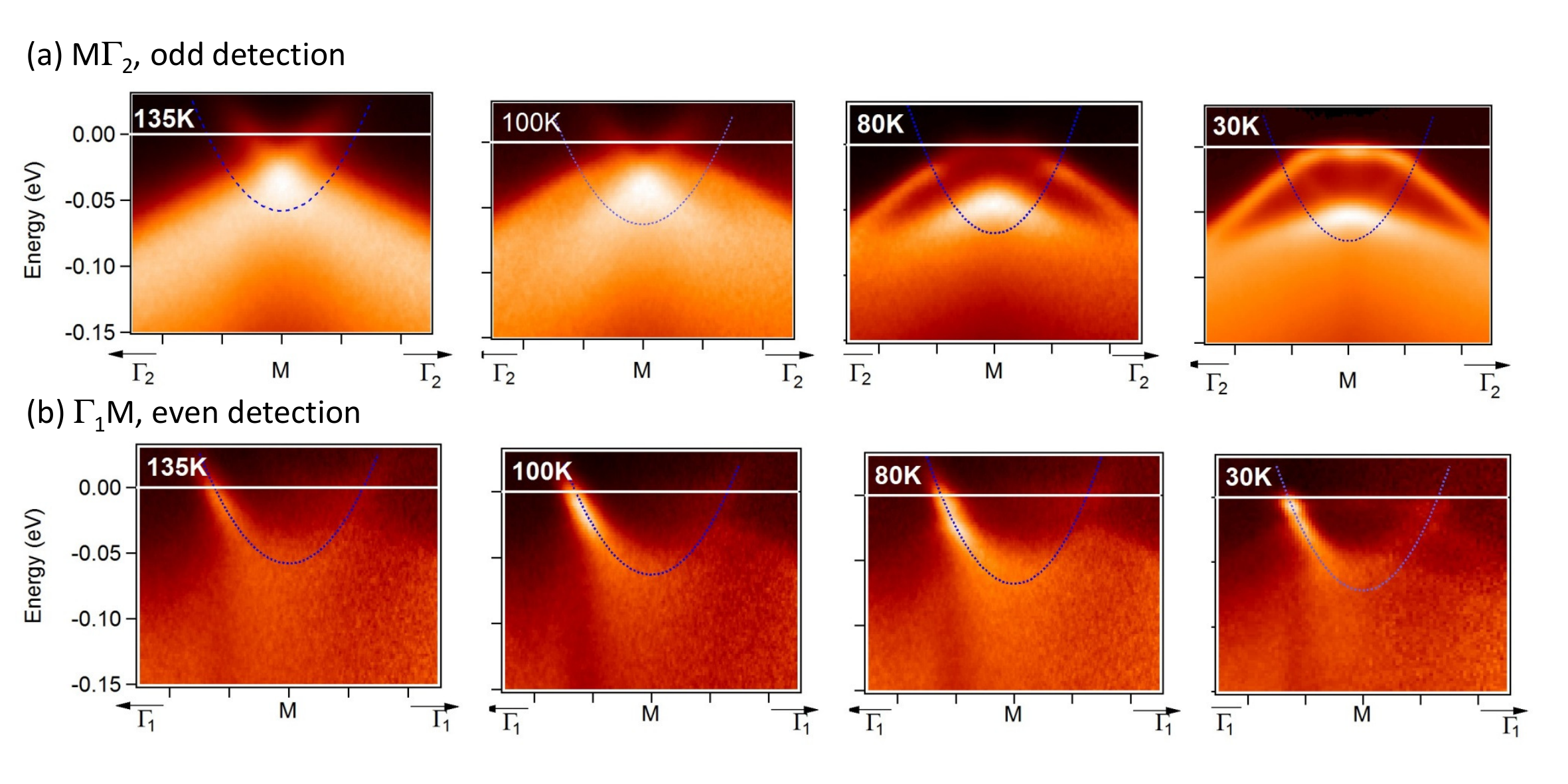}
 \caption{Temperature evolution at M point. (a) Energy-momentum plots around M in the same conditions as Fig.\ref{Fig_150K}(c), but with all EDC normalized to constant area, which emphasizes the saddle bands. (b) Energy-momentum plots around M in the same conditions as Fig.\ref{Fig_150K}(b), showing the $xy$ electron band. The fit of this band (dotted blue line) is reported in (a). $\Gamma_1$ and $\Gamma_2$ are the $\Gamma$ points at (0,0) and ($\pi$,$\pi$), respectively. } \label{Fig_TempSup}
 \end{figure*}

 \begin{figure}[tbp]
 \centering
 \includegraphics[width=0.45\textwidth]{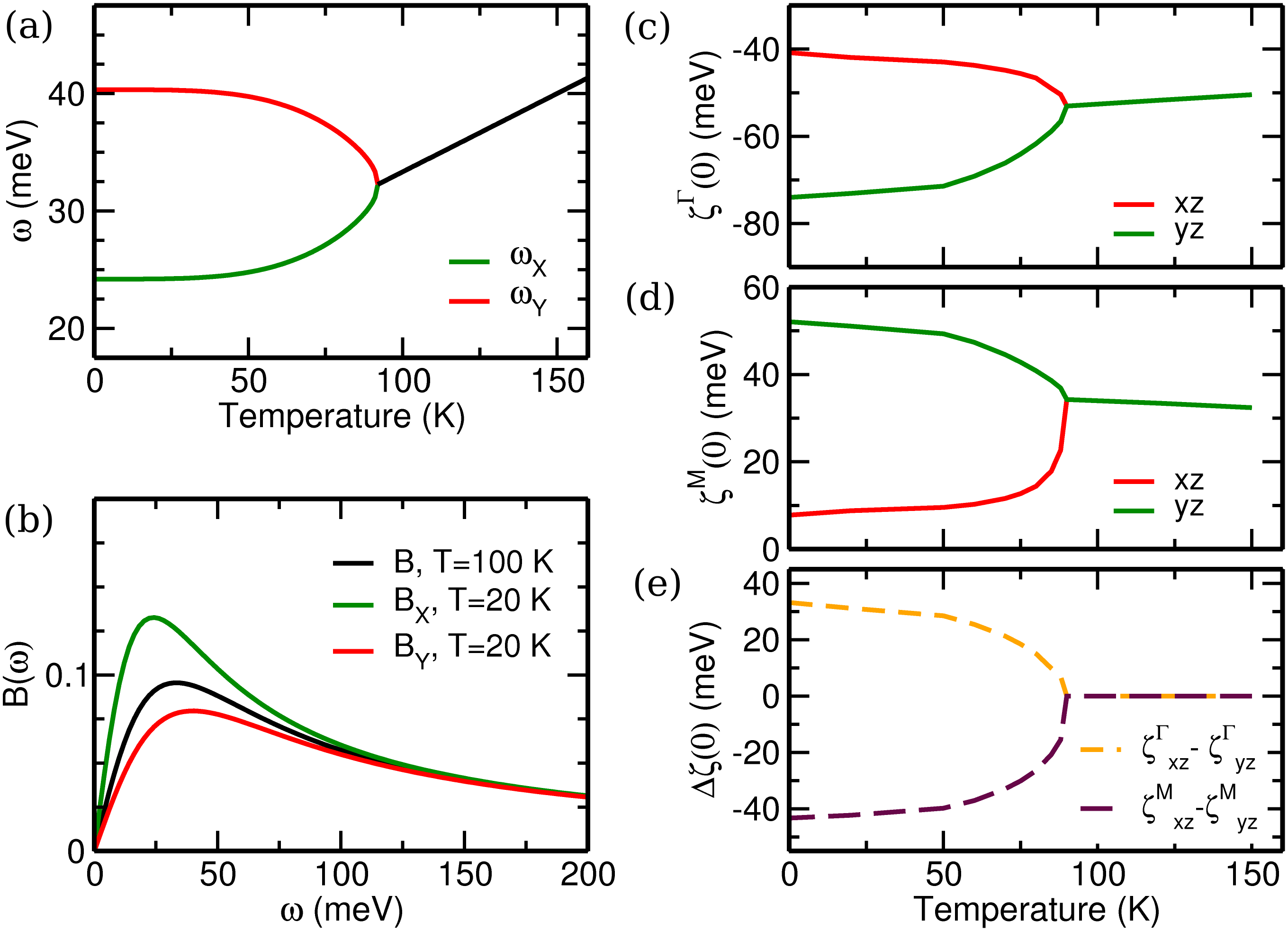}
 \caption{{\bf Evolution of the spin-fluctuation and nematic splitting with temperature}. (a) Temperature evolution of the SF energies $\o_{sf}^{X/Y}$ across the nematic transition. The SF propagator  \pref{bapp} becomes anisotropic below $T_S$, as shown in panel (b). (c-e) The anisotropy of the SF below $T_S$ induces a nematic splitting of the self-energy corrections for the $xz$ and $yz$ orbitals. Their real parts at zero frequency $\zeta$($\omega$=0) are shown in panel (c,d) for the  $\Gamma, M$ pockets. The resulting $xz/yz$ splitting below $T_S$ is shown in (d) .}
 \label{Fig_SF}
 \end{figure}

 \begin{figure*}[tbp]
 \centering
 \includegraphics[width=0.9\textwidth]{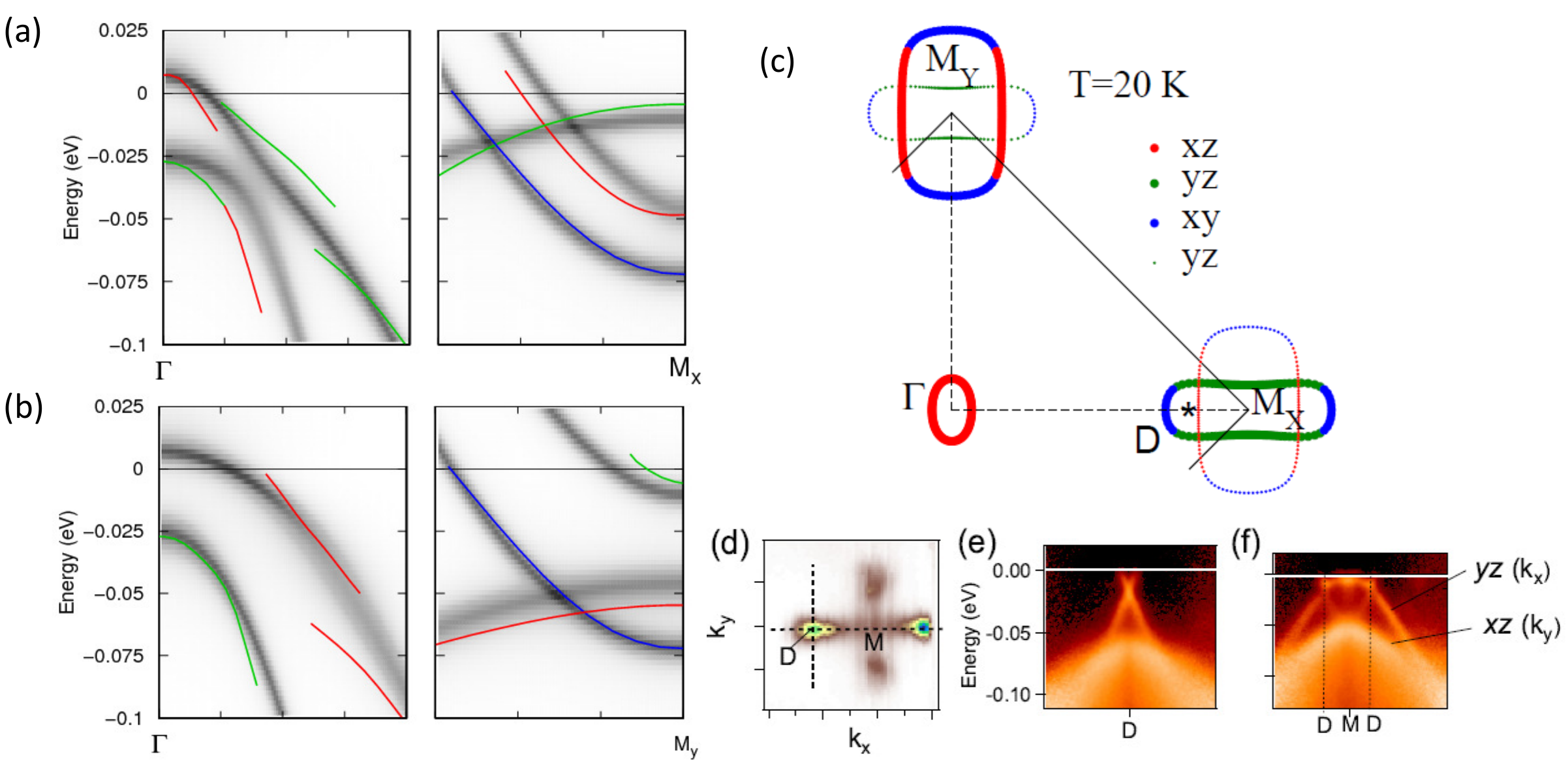}
 \caption{Electronic structure in the nematic state modeled in the spin-fluctuation scenario. Spectral functions of the renormalized bands at 20 K along $\G$M$_X$ (a) and $\G$M$_Y$ (b). Ticks along abscissa corresponds to 0.1 $\Gamma$M. Thin lines reproduce the experimental data detailed in Fig. \ref{Fig_20K}. The \lq\lq{}break\rq\rq{} at -60meV in the dispersion of the outer hole band is due the crossing of the $xy$ hole band. Along $k_x$ the main orbital character switches form $yz$ to $xz$ on the inner band and from $xz$ to $yz$ on the outer band, due to the crossing of the two bands hybridized through SOC. (c) Fermi Surface calculated for low temperatures. (d) Map of the ARPES spectral weight integrated around M at -15meV over 4meV and dispersions in the two perpendicular dispersion across the Dirac point D. (e-f) Energy-momentum plots in 2 perpendicular cuts across the Dirac point.}
 \label{Fig_Theory20K}
 \end{figure*}

In Fig. \ref{Fig_20K}(a-c), we show the same ARPES cuts as in Fig. \ref{Fig_150K}, but in the low temperature phase, at 20K. As for the high temperature case, we sketch all bands we observe by thin lines. They are also reported in Fig. \ref{Fig_20K}(d), along with dashed lines representing the bands at 150 K. 
The 50meV splitting at 20K between the saddle bands at M is the most dramatic feature of the nematic state and was already reported by many groups \cite{NakayamaPRL14,ShimojimaPRB14,WatsonPRB15,ZhangDingPRB15,ZhangShen15,Watson2016,BorCondMat16}. 

The correct assignment of the bands at M is crucial to identify which orbitals get split in the nematic phase. While the first measurements \cite{NakayamaPRL14,ShimojimaPRB14,WatsonPRB15,ZhangDingPRB15,ZhangShen15} interpreted the low-temperature data in terms of a 50 meV splitting between the $xz$ and $yz$ orbitals, recent reports \cite{Watson2016,BorCondMat16} suggested an alternative interpretation, with a large splitting between $xy$ at the bottom and a $xz/yz$ doublet at the top (whose degeneracy may be further removed by a small residual 15 meV splitting \cite{BorCondMat16}). As we clearly separate $xz/yz$ and $xy$ at all temperatures (see Fig. \ref{Fig_TempSup}), the distinction is easier in our case. Fig. \ref{Fig_TempSup}(b) shows that there is little change of the $xy$ electron band as a function of temperature, except for a small deepening of the band bottom at low temperatures, which we attribute to a reduction of the renormalization from 5 to 4.2 (see SI). Apart from this effect, the size of the pockets barely changes with temperature, showing that this band does not participate actively to the shrinking mechanism. We report the experimental shape of the $xy$ electron band on Fig. \ref{Fig_TempSup}(a) as a dotted line. As the bottom of the $xy$ electron band should be degenerate with the $xy$ saddle band (at least for high temperatures, more details are given in SI), the saddle band we observe moving down has to be $yz$, clearly supporting the existence of a large nematic splitting of the $xz/yz$ orbitals. The temperature evolution of the spectra at M is summarized in Fig.\ \ref{Fig_20K}(e-f), where we report the temperature dependence of the bottom of the bands at M and of their $k_F$. The three different $k_F$ at low temperatures is only consistent with the situation indicated in Fig. \ref{Fig_20K}(d). 

Around $\Gamma$ the orbital splitting is less obvious, since the experimental data of Fig. \ref{Fig_20K}(d) at 20K and 150K largely overlap. This effect led some authors to conclude that there is no or little change there \cite{ZhangDingPRB15,ZhangShen15}. However, there are some changes such as the appearance of a characteristic \lq\lq{}hat\rq\rq{} on top of the inner band in Fig. \ref{Fig_20K}(b). As was also concluded in ref. \cite{Suzuki15}, we will argue below that it is due to a SOC-avoided crossing of the two hole bands along $k_x$, which implies sizable shifts of the $xz/yz$ bands at $\Gamma$ also. 

\vspace{0.5cm}
The observed modifications of the Fermi surface can be very well understood within our orbital-selective spin-fluctuations scenario. Indeed the SF are expected to become inequivalent below $T_S$, as usually assumed for a spin-driven nematic transition\cite{FernandesNatPhys14}, making SF stronger at the wave-vector $\bQ_X$ where antiferromagnetic order usually develops at lower temperature. Even though long-range magnetic order is not observed in FeSe, SF have been detected \cite{Wang15,Rahn15,Shamoto15} and we assume that they become anisotropic below $T_S$, with a softening $\omega^{X}_{sf}<\omega_{sf}^Y$ that is the hallmark of having $\langle \bS\cdot \bS\rangle (\bQ_X)>\langle \bS\cdot \bS\rangle (\bQ_Y)$, as sketched in Fig.\ \ref{Fig_SF}(a-b). This has the immediate effect, from Eq.s\ \pref{corrx}-\pref{corry}, to split the self-energy corrections of the $xz$ and $yz$ orbitals, making in general their absolute values larger for $yz$. Taking into account  the different sign of the self-energy shifts \pref{sigmagx}-\pref{sigmagy} and \pref{sigmax}-\pref{sigmay}, respectively at the hole and electron pockets, one immediately finds that $\Delta\zeta^\G\equiv  \zeta^\G_{xz}-\zeta^\G_{yz}>0$ while  $\Delta\zeta^M<0$, as calculated in Fig.\ \ref{Fig_SF}(c-e). This is in agreement with previous experiments in detwinned samples \cite{ShimojimaPRB14,Suzuki15}. 

In Fig. \ref{Fig_Theory20K}, we report the theoretically computed spectral functions along $\Gamma$$M_X$ and $\Gamma$$M_Y$ and the FS. They are compared with our experimental data, where we have used the information from detwinned experiments \cite{ShimojimaPRB14,Suzuki15} and the above band assignment to determine which data lines correspond to measurements along $\G M_X$ or $\G M_Y$. Below $T_S$ the orbital-dependent shrinking induces an elliptical deformation of the hole pocket, which acquires mainly $xz$ character. At $M_X$, it shrinks the $yz$ orbital further, while at $M_Y$, the $xz$ orbital expands back. In order to reproduce quantitatively the experimental data, we computed the spectral functions of Fig. \ref{Fig_Theory20K} by using as a fitting parameter the splitting of the SF energies $\o_{sf}^{X/Y}$ (details are given in SI). A very good agreement is obtained with the paprameters of Fig. \ref{Fig_SF}. Note that the SF-induced splitting at $\Gamma$ (30meV) is comparable to the one at M (-40meV), evidencing that its effect on the overall band structure at $\G$ is less apparent only because it is hidden  by the strong SOC present there. We can then see from Fig.\ \ref{Fig_SF}b that the resulting anisotropy of the spin modes below $T_s$ is relatively weak, which is compatible with experiments in twinned samples \cite{Wang15,Rahn15,Shamoto15}. Nonetheless, the effects on the bands dispersions are quite strong. This is due to the particularly large particle-hole asymmetry of the bands in FeSe, resulting first from the strong renormalization inherent to FeSe and second from the high temperature shrinking.

The changes of the FS structure that we discussed so far are dictated by the real part of the SF-induced self-energy corrections. In addition, the imaginary part of the self-energy also becomes orbital dependent below $T_S$ and determines the lifetime of carriers in the nematic phase. Indeed, the dispersions shown in Fig. \ref{Fig_Theory20K}(e-f) suggest very different properties along $k_x$ and $k_y$ in the nematic state. Along $k_x$, a Dirac cone is formed between $xy$ and the $yz$ saddle band, just 15meV below $E_F$ \cite{TanFengPRB16,LiShen15}. The lines forming the Dirac cone are remarkably narrow, contrasting with the much broader lines belonging to the perpendicular domain, even for similar binding energies. Along $k_y$, the electron pocket, although enlarged, is made out of bands that appear very incoherent in ARPES. This coexistence of coherent and incoherent carriers is a very important and unusual characteristic of the metallic state of FeSe at low temperatures. 

\section {Conclusion}

The dichotomy between the orbital-driven and spin-driven  scenarios discussed so far for pnictides is connected to the difference between "hard" and "soft" possible realizations of electron nematicity. In the former case the nematic transition breaks explicitly a symmetry with the emergence of a finite electronic order parameter, that affects directly the single-particle electronic properties probed by ARPES. In the latter case instead the symmetry is broken only at the level of collective electronic fluctuations, that naturally affect collective two-particle properties, as probed e.g. by optical or Raman spectroscopy \cite{GallaisPRL13,HacklNatPhys16,BlumbergPRB16}. For this reason the strong modification of the electronic structure below $T_S$ seen by ARPES in FeSe triggered the idea \cite{KreiselCondMat15,LiangCondMat15} that a hard orbital order is needed to explain the modification of the electronic structure. However, while it is indeed consistent with the experimental observations, its  justification at microscopic level requires a specific fine-tuning of the interactions
\cite{ZhaiPRB09,JiangCondMat15,KontaniCondMat15}. On the other hand, in pnictides also collective fluctuations can affect the single-particle properties, and modify the Fermi surface \cite{OrtenziPRL09}. This mechanism, that relies on quite general conditions (the exchange of spin-fluctuations between hole and electron bands in pnictides) is already operative at high temperatures, and it naturally leads to the observed FS evolution in the 
nematic state. Thus, the orbital-selective spin-fluctuations scenario explains why a soft nematic transition can give rise, thanks to the strong spin-orbit entanglement, to an order-parameter like behavior of the electronic structure,  even in the absence of any hard symmetry breaking.  Besides reconciling different views, our results revise the standard paradigms for the understanding of hard and soft nematicity in the collective electronic behavior.

\vspace{1cm}
\noindent
\textbf{Acknowledgements}
V.B. and J.M. acknowledge financial support from the ANR \lq\lq{}PNICTIDES\rq\rq{}. L.B. acknowledges financial support by Italian MIUR under projects FIRB-HybridNanoDev-RBFR1236VV, PRINRIDEIRON-2012X3YFZ2 and Premiali-2012 ABNANOTECH. B.V acknowledges funding from Ministerio de Econom\'ia y Competitividad (Spain) via Grants No.FIS2014-53219-P, No. FIS2011-29689   and Fundaci\'on Ram\'on Areces.

\bibliography{FeSe_Bib_LB}

\begin{thebibliography}{10}%
\makeatletter
\providecommand \@ifxundefined [1]{%
 \ifx #1\undefined \expandafter \@firstoftwo
 \else \expandafter \@secondoftwo
\fi
}%
\providecommand \@ifnum [1]{%
 \ifnum #1\expandafter \@firstoftwo
 \else \expandafter \@secondoftwo
\fi
}%
\providecommand \enquote [1]{``#1''}%
\providecommand \bibnamefont  [1]{#1}%
\providecommand \bibfnamefont [1]{#1}%
\providecommand \citenamefont [1]{#1}%
\providecommand\href[0]{\@sanitize\@href}%
\providecommand\@href[1]{\endgroup\@@startlink{#1}\endgroup\@@href}%
\providecommand\@@href[1]{#1\@@endlink}%
\providecommand \@sanitize [0]{\begingroup\catcode`\&12\catcode`\#12\relax}%
\@ifxundefined \pdfoutput {\@firstoftwo}{%
 \@ifnum{\z@=\pdfoutput}{\@firstoftwo}{\@secondoftwo}%
}{%
 \providecommand\@@startlink[1]{\leavevmode\special{html:<a href="#1">}}%
 \providecommand\@@endlink[0]{\special{html:</a>}}%
}{%
 \providecommand\@@startlink[1]{%
  \leavevmode
  \pdfstartlink
   attr{/Border[0 0 1 ]/H/I/C[0 1 1]}%
   user{/Subtype/Link/A<</Type/Action/S/URI/URI(#1)>>}%
  \relax
 }%
 \providecommand\@@endlink[0]{\pdfendlink}%
}%
\providecommand \url  [0]{\begingroup\@sanitize \@url }%
\providecommand \@url [1]{\endgroup\@href {#1}{\urlprefix}}%
\providecommand \urlprefix [0]{URL }%
\providecommand \Eprint[0]{\href }%
\@ifxundefined \urlstyle {%
  \providecommand \doi [1]{doi:\discretionary{}{}{}#1}%
}{%
  \providecommand \doi [0]{doi:\discretionary{}{}{}\begingroup
  \urlstyle{rm}\Url }%
}%
\providecommand \doibase [0]{http://dx.doi.org/}%
\providecommand \Doi[1]{\href{\doibase#1}}%
\providecommand \bibAnnote [3]{%
  \BibitemShut{#1}%
  \begin{quotation}\noindent
    \textsc{Key:}\ #2\\\textsc{Annotation:}\ #3%
  \end{quotation}%
}%
\providecommand \bibAnnoteFile [2]{%
  \IfFileExists{#2}{\bibAnnote {#1} {#2} {\input{#2}}}{}%
}%
\providecommand \typeout [0]{\immediate \write \m@ne }%
\providecommand \selectlanguage [0]{\@gobble}%
\providecommand \bibinfo [0]{\@secondoftwo}%
\providecommand \bibfield [0]{\@secondoftwo}%
\providecommand \translation [1]{[#1]}%
\providecommand \BibitemOpen[0]{}%
\providecommand \bibitemStop [0]{}%
\providecommand \bibitemNoStop [0]{.\EOS\space}%
\providecommand \EOS [0]{\spacefactor3000\relax}%
\providecommand \BibitemShut [1]{\csname bibitem#1\endcsname}%
\bibitem{FradkinReview2010}%
  \BibitemOpen
  \bibfield{author}{%
  \bibinfo {author} {\bibfnamefont{E.}~\bibnamefont{Fradkin}}, \bibinfo
  {author} {\bibfnamefont{S.}~\bibnamefont{Kivelson}}, \bibinfo {author}
  {\bibfnamefont{M.}~\bibnamefont{Lawler}}, \bibinfo {author}
  {\bibfnamefont{J.}~\bibnamefont{Eisenstein}},\ and\ \bibinfo {author}
  {\bibfnamefont{A.}~\bibnamefont{Mackenzie}},\ }%
  \bibfield{journal}{%
  \bibinfo {journal} {Annual Review of Condensed Matter Physics}\ }%
  \textbf{\bibinfo {volume} {1}},\ \bibinfo {pages} {153} (\bibinfo {year}
  {2010})%
  \bibAnnoteFile{NoStop}{FradkinReview2010}%
\bibitem{FisherRPP2011}%
  \BibitemOpen
  \bibfield{author}{%
  \bibinfo {author} {\bibfnamefont{I.~R.}\ \bibnamefont{Fisher}}, \bibinfo
  {author} {\bibfnamefont{L.}~\bibnamefont{Degiorgi}},\ and\ \bibinfo {author}
  {\bibfnamefont{Z.}~\bibnamefont{Shen}},\ }%
  \bibfield{journal}{%
  \bibinfo {journal} {Rep. Prog. Phys.}\ }%
  \textbf{\bibinfo {volume} {74}},\ \bibinfo {pages} {124506} (\bibinfo {year}
  {2011})%
  \bibAnnoteFile{NoStop}{FisherRPP2011}%
\bibitem{FernandesNatPhys14}%
  \BibitemOpen
  \bibfield{author}{%
  \bibinfo {author} {\bibfnamefont{R.~M.}\ \bibnamefont{Fernandes}}, \bibinfo
  {author} {\bibfnamefont{A.~V.}\ \bibnamefont{Chubukov}},\ and\ \bibinfo
  {author} {\bibfnamefont{J.}~\bibnamefont{Schmalian}},\ }%
  \bibfield{journal}{%
  \bibinfo {journal} {Nature Physics}\ }%
  \textbf{\bibinfo {volume} {10}},\ \bibinfo {pages} {98} (\bibinfo {year}
  {2014})%
  \bibAnnoteFile{NoStop}{FernandesNatPhys14}%
\bibitem{GallaisReview15}%
  \BibitemOpen
  \bibfield{author}{%
  \bibinfo {author} {\bibfnamefont{Y.}~\bibnamefont{Gallais}}\ and\ \bibinfo
  {author} {\bibfnamefont{I.}~\bibnamefont{Paul}},\ }%
  \bibfield{journal}{%
  \bibinfo {journal} {Comptes rendus Phys.}\ }%
  \textbf{\bibinfo {volume} {17}},\ \bibinfo {pages} {113} (\bibinfo {year}
  {2016})%
  \bibAnnoteFile{NoStop}{GallaisReview15}%
\bibitem{BohmerPRB13}%
  \BibitemOpen
  \bibfield{author}{%
  \bibinfo {author} {\bibfnamefont{A.~E.}\ \bibnamefont{B\"ohmer}}, \bibinfo
  {author} {\bibfnamefont{F.}~\bibnamefont{Hardy}}, \bibinfo {author}
  {\bibfnamefont{F.}~\bibnamefont{Eilers}}, \bibinfo {author}
  {\bibfnamefont{D.}~\bibnamefont{Ernst}}, \bibinfo {author}
  {\bibfnamefont{P.}~\bibnamefont{Adelmann}}, \bibinfo {author}
  {\bibfnamefont{P.}~\bibnamefont{Schweiss}}, \bibinfo {author}
  {\bibfnamefont{T.}~\bibnamefont{Wolf}},\ and\ \bibinfo {author}
  {\bibfnamefont{C.}~\bibnamefont{Meingast}},\ }%
  \bibfield{journal}{%
  \bibinfo {journal} {Phys. Rev. B}\ }%
  \textbf{\bibinfo {volume} {87}},\ \bibinfo {pages} {180505} (\bibinfo {year}
  {2013})%
  \bibAnnoteFile{NoStop}{BohmerPRB13}%
\bibitem{HsuPNAS08}%
  \BibitemOpen
  \bibfield{author}{%
  \bibinfo {author} {\bibfnamefont{F.}~\bibnamefont{Hsu}} \emph{et~al.},\ }%
  \bibfield{journal}{%
  \bibinfo {journal} {PNAS}\ }%
  \textbf{\bibinfo {volume} {105}},\ \bibinfo {pages} {14262} (\bibinfo {year}
  {2008})%
  \bibAnnoteFile{NoStop}{HsuPNAS08}%
\bibitem{Tanatarcm15}%
  \BibitemOpen
  \bibfield{author}{%
  \bibinfo {author} {\bibfnamefont{M.}~\bibnamefont{Tanatar}} \emph{et~al.},\
  }%
  \bibinfo {journal} {arXiv:1511.04757}%
  \bibAnnoteFile{NoStop}{Tanatarcm15}%
\bibitem{ShimojimaPRB14}%
  \BibitemOpen
\bibfield{journal}{%
    }%
  \bibfield{author}{%
  \bibinfo {author} {\bibfnamefont{T.}~\bibnamefont{Shimojima}}, \bibinfo
  {author} {\bibfnamefont{Y.}~\bibnamefont{Suzuki}}, \bibinfo {author}
  {\bibfnamefont{T.}~\bibnamefont{Sonobe}}, \bibinfo {author}
  {\bibfnamefont{A.}~\bibnamefont{Nakamura}}, \bibinfo {author}
  {\bibfnamefont{M.}~\bibnamefont{Sakano}}, \bibinfo {author}
  {\bibfnamefont{J.}~\bibnamefont{Omachi}}, \bibinfo {author}
  {\bibfnamefont{K.}~\bibnamefont{Yoshioka}}, \bibinfo {author}
  {\bibfnamefont{M.}~\bibnamefont{Kuwata-Gonokami}}, \bibinfo {author}
  {\bibfnamefont{K.}~\bibnamefont{Ono}}, \bibinfo {author}
  {\bibfnamefont{H.}~\bibnamefont{Kumigashira}}, \bibinfo {author}
  {\bibfnamefont{A.~E.}\ \bibnamefont{Bohmer}}, \bibinfo {author}
  {\bibfnamefont{F.}~\bibnamefont{Hardy}}, \bibinfo {author}
  {\bibfnamefont{T.}~\bibnamefont{Wolf}}, \bibinfo {author}
  {\bibfnamefont{C.}~\bibnamefont{Meingast}}, \bibinfo {author}
  {\bibfnamefont{H.~v.}\ \bibnamefont{Lohneysen}}, \bibinfo {author}
  {\bibfnamefont{H.}~\bibnamefont{Ikeda}},\ and\ \bibinfo {author}
  {\bibfnamefont{K.}~\bibnamefont{Ishizaka}},\ }%
  \bibfield{journal}{%
  \bibinfo {journal} {Phys. Rev. B}\ }%
  \textbf{\bibinfo {volume} {90}},\ \bibinfo {pages} {121111} (\bibinfo {year}
  {2014})%
  \bibAnnoteFile{NoStop}{ShimojimaPRB14}%
\bibitem{NakayamaPRL14}%
  \BibitemOpen
  \bibfield{author}{%
  \bibinfo {author} {\bibfnamefont{K.}~\bibnamefont{Nakayama}}, \bibinfo
  {author} {\bibfnamefont{Y.}~\bibnamefont{Miyata}}, \bibinfo {author}
  {\bibfnamefont{G.~N.}\ \bibnamefont{Phan}}, \bibinfo {author}
  {\bibfnamefont{T.}~\bibnamefont{Sato}}, \bibinfo {author}
  {\bibfnamefont{Y.}~\bibnamefont{Tanabe}}, \bibinfo {author}
  {\bibfnamefont{T.}~\bibnamefont{Urata}}, \bibinfo {author}
  {\bibfnamefont{K.}~\bibnamefont{Tanigaki}},\ and\ \bibinfo {author}
  {\bibfnamefont{T.}~\bibnamefont{Takahashi}},\ }%
  \bibfield{journal}{%
  \bibinfo {journal} {Phys. Rev. Lett.}\ }%
  \textbf{\bibinfo {volume} {113}},\ \bibinfo {pages} {237001} (\bibinfo {year}
  {2014})%
  \bibAnnoteFile{NoStop}{NakayamaPRL14}%
\bibitem{WatsonPRB15}%
  \BibitemOpen
  \bibfield{author}{%
  \bibinfo {author} {\bibfnamefont{M.~D.}\ \bibnamefont{Watson}}, \bibinfo
  {author} {\bibfnamefont{T.~K.}\ \bibnamefont{Kim}}, \bibinfo {author}
  {\bibfnamefont{A.~A.}\ \bibnamefont{Haghighirad}}, \bibinfo {author}
  {\bibfnamefont{N.~R.}\ \bibnamefont{Davies}}, \bibinfo {author}
  {\bibfnamefont{A.}~\bibnamefont{McCollam}}, \bibinfo {author}
  {\bibfnamefont{A.}~\bibnamefont{Narayanan}}, \bibinfo {author}
  {\bibfnamefont{S.~F.}\ \bibnamefont{Blake}}, \bibinfo {author}
  {\bibfnamefont{Y.~L.}\ \bibnamefont{Chen}}, \bibinfo {author}
  {\bibfnamefont{S.}~\bibnamefont{Ghannadzadeh}}, \bibinfo {author}
  {\bibfnamefont{A.~J.}\ \bibnamefont{Schofield}}, \bibinfo {author}
  {\bibfnamefont{M.}~\bibnamefont{Hoesch}}, \bibinfo {author}
  {\bibfnamefont{C.}~\bibnamefont{Meingast}}, \bibinfo {author}
  {\bibfnamefont{T.}~\bibnamefont{Wolf}},\ and\ \bibinfo {author}
  {\bibfnamefont{A.~I.}\ \bibnamefont{Coldea}},\ }%
  \bibfield{journal}{%
  \bibinfo {journal} {Phys. Rev. B}\ }%
  \textbf{\bibinfo {volume} {91}},\ \bibinfo {pages} {155106} (\bibinfo {year}
  {2015})%
  \bibAnnoteFile{NoStop}{WatsonPRB15}%
\bibitem{ZhangDingPRB15}%
  \BibitemOpen
  \bibfield{author}{%
  \bibinfo {author} {\bibfnamefont{P.}~\bibnamefont{Zhang}}, \bibinfo {author}
  {\bibfnamefont{T.}~\bibnamefont{Qian}}, \bibinfo {author}
  {\bibfnamefont{P.}~\bibnamefont{Richard}}, \bibinfo {author}
  {\bibfnamefont{X.~P.}\ \bibnamefont{Wang}}, \bibinfo {author}
  {\bibfnamefont{H.}~\bibnamefont{Miao}}, \bibinfo {author}
  {\bibfnamefont{B.~Q.}\ \bibnamefont{Lv}}, \bibinfo {author}
  {\bibfnamefont{B.~B.}\ \bibnamefont{Fu}}, \bibinfo {author}
  {\bibfnamefont{T.}~\bibnamefont{Wolf}}, \bibinfo {author}
  {\bibfnamefont{C.}~\bibnamefont{Meingast}}, \bibinfo {author}
  {\bibfnamefont{X.~X.}\ \bibnamefont{Wu}}, \bibinfo {author}
  {\bibfnamefont{Z.~Q.}\ \bibnamefont{Wang}}, \bibinfo {author}
  {\bibfnamefont{J.~P.}\ \bibnamefont{Hu}},\ and\ \bibinfo {author}
  {\bibfnamefont{H.}~\bibnamefont{Ding}},\ }%
  \bibfield{journal}{%
  \bibinfo {journal} {Phys. Rev. B}\ }%
  \textbf{\bibinfo {volume} {91}},\ \bibinfo {pages} {214503} (\bibinfo {year}
  {2015})%
  \bibAnnoteFile{NoStop}{ZhangDingPRB15}%
\bibitem{ZhangShen15}%
  \BibitemOpen
  \bibfield{author}{%
  \bibinfo {author} {\bibfnamefont{Y.}~\bibnamefont{Zhang}} \emph{et~al.},\ }%
  \bibinfo {journal} {arXiv:1503.01556}%
  \bibAnnoteFile{NoStop}{ZhangShen15}%
\bibitem{Suzuki15}%
  \BibitemOpen
\bibfield{journal}{%
    }%
  \bibfield{author}{%
  \bibinfo {author} {\bibfnamefont{Y.}~\bibnamefont{Suzuki}}, \bibinfo {author}
  {\bibfnamefont{T.}~\bibnamefont{Shimojima}}, \bibinfo {author}
  {\bibfnamefont{T.}~\bibnamefont{Sonobe}}, \bibinfo {author}
  {\bibfnamefont{A.}~\bibnamefont{Nakamura}}, \bibinfo {author}
  {\bibfnamefont{M.}~\bibnamefont{Sakano}}, \bibinfo {author}
  {\bibfnamefont{H.}~\bibnamefont{Tsuji}}, \bibinfo {author}
  {\bibfnamefont{J.}~\bibnamefont{Omachi}}, \bibinfo {author}
  {\bibfnamefont{K.}~\bibnamefont{Yoshioka}}, \bibinfo {author}
  {\bibfnamefont{M.}~\bibnamefont{Kuwata-Gonokami}}, \bibinfo {author}
  {\bibfnamefont{T.}~\bibnamefont{Watashige}}, \bibinfo {author}
  {\bibfnamefont{R.}~\bibnamefont{Kobayashi}}, \bibinfo {author}
  {\bibfnamefont{S.}~\bibnamefont{Kasahara}}, \bibinfo {author}
  {\bibfnamefont{T.}~\bibnamefont{Shibauchi}}, \bibinfo {author}
  {\bibfnamefont{Y.}~\bibnamefont{Matsuda}}, \bibinfo {author}
  {\bibfnamefont{Y.}~\bibnamefont{Yamakawa}}, \bibinfo {author}
  {\bibfnamefont{H.}~\bibnamefont{Kontani}},\ and\ \bibinfo {author}
  {\bibfnamefont{K.}~\bibnamefont{Ishizaka}},\ }%
  \bibfield{journal}{%
  \bibinfo {journal} {Phys. Rev. B}\ }%
  \textbf{\bibinfo {volume} {92}},\ \bibinfo {pages} {205117} (\bibinfo {year}
  {2015})%
  \bibAnnoteFile{NoStop}{Suzuki15}%
\bibitem{LvPRB09}%
  \BibitemOpen
  \bibfield{author}{%
  \bibinfo {author} {\bibfnamefont{W.}~\bibnamefont{Lv}}, \bibinfo {author}
  {\bibfnamefont{J.}~\bibnamefont{Wu}},\ and\ \bibinfo {author}
  {\bibfnamefont{P.}~\bibnamefont{Phillips}},\ }%
  \bibfield{journal}{%
  \bibinfo {journal} {Phys. Rev. B}\ }%
  \textbf{\bibinfo {volume} {80}},\ \bibinfo {pages} {224506} (\bibinfo {month}
  {Jan}\ \bibinfo {year} {2009})%
  \bibAnnoteFile{NoStop}{LvPRB09}%
\bibitem{SuJcondMat15}%
  \BibitemOpen
  \bibfield{author}{%
  \bibinfo {author} {\bibfnamefont{Y.}~\bibnamefont{Su}}, \bibinfo {author}
  {\bibfnamefont{H.}~\bibnamefont{Liao}},\ and\ \bibinfo {author}
  {\bibfnamefont{T.}~\bibnamefont{Li}},\ }%
  \bibfield{journal}{%
  \bibinfo {journal} {Journal of Physics: Condensed Matter}\ }%
  \textbf{\bibinfo {volume} {27}},\ \bibinfo {pages} {105702} (\bibinfo {year}
  {2015})%
  \bibAnnoteFile{NoStop}{SuJcondMat15}%
\bibitem{MukherjeePRL15}%
  \BibitemOpen
  \bibfield{author}{%
  \bibinfo {author} {\bibfnamefont{S.}~\bibnamefont{Mukherjee}}, \bibinfo
  {author} {\bibfnamefont{A.}~\bibnamefont{Kreisel}}, \bibinfo {author}
  {\bibfnamefont{P.~J.}\ \bibnamefont{Hirschfeld}},\ and\ \bibinfo {author}
  {\bibfnamefont{B.~M.}\ \bibnamefont{Andersen}},\ }%
  \bibfield{journal}{%
  \bibinfo {journal} {Phys. Rev. Lett.}\ }%
  \textbf{\bibinfo {volume} {115}},\ \bibinfo {pages} {026402} (\bibinfo {year}
  {2015})%
  \bibAnnoteFile{NoStop}{MukherjeePRL15}%
\bibitem{JiangCondMat15}%
  \BibitemOpen
  \bibfield{author}{%
  \bibinfo {author} {\bibfnamefont{K.}~\bibnamefont{Jiang}}, \bibinfo {author}
  {\bibfnamefont{J.}~\bibnamefont{Hu}}, \bibinfo {author}
  {\bibfnamefont{H.}~\bibnamefont{Ding}},\ and\ \bibinfo {author}
  {\bibfnamefont{Z.}~\bibnamefont{Wang}},\ }%
  \bibfield{journal}{%
  \bibinfo {journal} {Phys. Rev. B}\ }%
  \textbf{\bibinfo {volume} {93}},\ \bibinfo {pages} {115138} (\bibinfo {year}
  {2016})%
  \bibAnnoteFile{NoStop}{JiangCondMat15}%
\bibitem{BaekNatMat14}%
  \BibitemOpen
  \bibfield{author}{%
  \bibinfo {author} {\bibfnamefont{S.}~\bibnamefont{Baek}}, \bibinfo {author}
  {\bibfnamefont{D.}~\bibnamefont{Efremov}}, \bibinfo {author}
  {\bibfnamefont{J.~M.}\ \bibnamefont{Ok}}, \bibinfo {author}
  {\bibfnamefont{J.~S.}\ \bibnamefont{Kim}}, \bibinfo {author}
  {\bibfnamefont{J.}~\bibnamefont{van~den Brink}},\ and\ \bibinfo {author}
  {\bibfnamefont{B.}~\bibnamefont{Buchner}},\ }%
  \bibfield{journal}{%
  \bibinfo {journal} {Nature Materials}\ }%
  \textbf{\bibinfo {volume} {14}},\ \bibinfo {pages} {210} (\bibinfo {year}
  {2014})%
  \bibAnnoteFile{NoStop}{BaekNatMat14}%
\bibitem{BohmerPRL15}%
  \BibitemOpen
  \bibfield{author}{%
  \bibinfo {author} {\bibfnamefont{A.}~\bibnamefont{Bohmer}} \emph{et~al.},\ }%
  \bibfield{journal}{%
  \bibinfo {journal} {Physical Review Letters}\ }%
  \textbf{\bibinfo {volume} {114}},\ \bibinfo {pages} {27001} (\bibinfo {year}
  {2015})%
  \bibAnnoteFile{NoStop}{BohmerPRL15}%
\bibitem{Wang15}%
  \BibitemOpen
  \bibfield{author}{%
  \bibinfo {author} {\bibfnamefont{Q.}~\bibnamefont{Wang}} \emph{et~al.},\ }%
  \bibfield{journal}{%
  \bibinfo {journal} {Nature Materials}\ }%
  \textbf{\bibinfo {volume} {15}},\ \bibinfo {pages} {159} (\bibinfo {year}
  {2016})%
  \bibAnnoteFile{NoStop}{Wang15}%
\bibitem{Rahn15}%
  \BibitemOpen
  \bibfield{author}{%
  \bibinfo {author} {\bibfnamefont{M.~C.}\ \bibnamefont{Rahn}}, \bibinfo
  {author} {\bibfnamefont{R.~A.}\ \bibnamefont{Ewings}}, \bibinfo {author}
  {\bibfnamefont{S.~J.}\ \bibnamefont{Sedlmaier}}, \bibinfo {author}
  {\bibfnamefont{S.~J.}\ \bibnamefont{Clarke}},\ and\ \bibinfo {author}
  {\bibfnamefont{A.~T.}\ \bibnamefont{Boothroyd}},\ }%
  \bibfield{journal}{%
  \bibinfo {journal} {Phys. Rev. B}\ }%
  \textbf{\bibinfo {volume} {91}},\ \bibinfo {pages} {180501} (\bibinfo {year}
  {2015})%
  \bibAnnoteFile{NoStop}{Rahn15}%
\bibitem{Shamoto15}%
  \BibitemOpen
  \bibfield{author}{%
  \bibinfo {author} {\bibfnamefont{S.}~\bibnamefont{Shamoto}} \emph{et~al.},\
  }%
  \bibinfo {journal} {arXiv:1511.04267}%
  \bibAnnoteFile{NoStop}{Shamoto15}%
\bibitem{VasilievCondMat16}%
  \BibitemOpen
\bibfield{journal}{%
    }%
  \bibfield{author}{%
  \bibinfo {author} {\bibfnamefont{C.}~\bibnamefont{Lo}} \emph{et~al.},\ }%
  \bibinfo {journal} {arXiv:1603.08710}%
  \bibAnnoteFile{NoStop}{VasilievCondMat16}%
\bibitem{Glasbrenner15}%
  \BibitemOpen
\bibfield{journal}{%
    }%
  \bibfield{author}{%
  \bibinfo {author} {\bibfnamefont{J.~K.}\ \bibnamefont{Glasbrenner}}
  \emph{et~al.},\ }%
  \bibfield{journal}{%
  \bibinfo {journal} {Nature Physics}\ }%
  \textbf{\bibinfo {volume} {11}},\ \bibinfo {pages} {953} (\bibinfo {year}
  {2015})%
  \bibAnnoteFile{NoStop}{Glasbrenner15}%
\bibitem{coldeaPRL08}%
  \BibitemOpen
  \bibfield{author}{%
  \bibinfo {author} {\bibfnamefont{A.~I.}\ \bibnamefont{Coldea}}, \bibinfo
  {author} {\bibfnamefont{J.~D.}\ \bibnamefont{Fletcher}}, \bibinfo {author}
  {\bibfnamefont{A.}~\bibnamefont{Carrington}}, \bibinfo {author}
  {\bibfnamefont{J.~G.}\ \bibnamefont{Analytis}}, \bibinfo {author}
  {\bibfnamefont{A.~F.}\ \bibnamefont{Bangura}}, \bibinfo {author}
  {\bibfnamefont{J.}~\bibnamefont{Chu}}, \bibinfo {author}
  {\bibfnamefont{A.~S.}\ \bibnamefont{Erickson}}, \bibinfo {author}
  {\bibfnamefont{I.~R.}\ \bibnamefont{Fisher}}, \bibinfo {author}
  {\bibfnamefont{N.~E.}\ \bibnamefont{Hussey}},\ and\ \bibinfo {author}
  {\bibfnamefont{R.~D.}\ \bibnamefont{McDonald}},\ }%
  \bibfield{journal}{%
  \bibinfo {journal} {Phys. Rev. Lett.}\ }%
  \textbf{\bibinfo {volume} {101}},\ \bibinfo {pages} {216402} (\bibinfo {year}
  {2008})%
  \bibAnnoteFile{NoStop}{coldeaPRL08}%
\bibitem{BrouetPRL13}%
  \BibitemOpen
  \bibfield{author}{%
  \bibinfo {author} {\bibfnamefont{V.}~\bibnamefont{Brouet}}, \bibinfo {author}
  {\bibfnamefont{P.-H.}\ \bibnamefont{Lin}}, \bibinfo {author}
  {\bibfnamefont{Y.}~\bibnamefont{Texier}}, \bibinfo {author}
  {\bibfnamefont{J.}~\bibnamefont{Bobroff}}, \bibinfo {author}
  {\bibfnamefont{A.}~\bibnamefont{Taleb-Ibrahimi}}, \bibinfo {author}
  {\bibfnamefont{P.}~\bibnamefont{Le~F\`evre}}, \bibinfo {author}
  {\bibfnamefont{F.}~\bibnamefont{Bertran}}, \bibinfo {author}
  {\bibfnamefont{M.}~\bibnamefont{Casula}}, \bibinfo {author}
  {\bibfnamefont{P.}~\bibnamefont{Werner}}, \bibinfo {author}
  {\bibfnamefont{S.}~\bibnamefont{Biermann}}, \bibinfo {author}
  {\bibfnamefont{F.}~\bibnamefont{Rullier-Albenque}}, \bibinfo {author}
  {\bibfnamefont{A.}~\bibnamefont{Forget}},\ and\ \bibinfo {author}
  {\bibfnamefont{D.}~\bibnamefont{Colson}},\ }%
  \bibfield{journal}{%
  \bibinfo {journal} {Phys. Rev. Lett.}\ }%
  \textbf{\bibinfo {volume} {110}},\ \bibinfo {pages} {167002} (\bibinfo {year}
  {2013})%
  \bibAnnoteFile{NoStop}{BrouetPRL13}%
\bibitem{Demedici14}%
  \BibitemOpen
  \bibfield{author}{%
  \bibinfo {author} {\bibfnamefont{L.}~\bibnamefont{de' Medici}}, \bibinfo
  {author} {\bibfnamefont{G.}~\bibnamefont{Giovannetti}},\ and\ \bibinfo
  {author} {\bibfnamefont{M.}~\bibnamefont{Capone}},\ }%
  \bibfield{journal}{%
  \bibinfo {journal} {Phys. Rev. Lett.}\ }%
  \textbf{\bibinfo {volume} {112}},\ \bibinfo {pages} {177001} (\bibinfo {year}
  {2014})%
  \bibAnnoteFile{NoStop}{Demedici14}%
\bibitem{AichhornPRBFeSe}%
  \BibitemOpen
  \bibfield{author}{%
  \bibinfo {author} {\bibfnamefont{M.}~\bibnamefont{Aichhorn}}, \bibinfo
  {author} {\bibfnamefont{S.}~\bibnamefont{Biermann}}, \bibinfo {author}
  {\bibfnamefont{T.}~\bibnamefont{Miyake}}, \bibinfo {author}
  {\bibfnamefont{A.}~\bibnamefont{Georges}},\ and\ \bibinfo {author}
  {\bibfnamefont{M.}~\bibnamefont{Imada}},\ }%
  \bibfield{journal}{%
  \bibinfo {journal} {Phys. Rev. B}\ }%
  \textbf{\bibinfo {volume} {82}},\ \bibinfo {pages} {064504} (\bibinfo {year}
  {2010})%
  \bibAnnoteFile{NoStop}{AichhornPRBFeSe}%
\bibitem{YinPRB12}%
  \BibitemOpen
  \bibfield{author}{%
  \bibinfo {author} {\bibfnamefont{Z.~P.}\ \bibnamefont{Yin}}, \bibinfo
  {author} {\bibfnamefont{K.}~\bibnamefont{Haule}},\ and\ \bibinfo {author}
  {\bibfnamefont{G.}~\bibnamefont{Kotliar}},\ }%
  \bibfield{journal}{%
  \bibinfo {journal} {Phys. Rev. B}\ }%
  \textbf{\bibinfo {volume} {86}},\ \bibinfo {pages} {195141} (\bibinfo {year}
  {2012})%
  \bibAnnoteFile{NoStop}{YinPRB12}%
\bibitem{OrtenziPRL09}%
  \BibitemOpen
  \bibfield{author}{%
  \bibinfo {author} {\bibfnamefont{L.}~\bibnamefont{Ortenzi}}, \bibinfo
  {author} {\bibfnamefont{E.}~\bibnamefont{Cappelluti}}, \bibinfo {author}
  {\bibfnamefont{L.}~\bibnamefont{Benfatto}},\ and\ \bibinfo {author}
  {\bibfnamefont{L.}~\bibnamefont{Pietronero}},\ }%
  \bibfield{journal}{%
  \bibinfo {journal} {Phys. Rev. Lett.}\ }%
  \textbf{\bibinfo {volume} {103}},\ \bibinfo {pages} {046404} (\bibinfo {year}
  {2009})%
  \bibAnnoteFile{NoStop}{OrtenziPRL09}%
\bibitem{BenfattoPRB11}%
  \BibitemOpen
  \bibfield{author}{%
  \bibinfo {author} {\bibfnamefont{L.}~\bibnamefont{Benfatto}}\ and\ \bibinfo
  {author} {\bibfnamefont{E.}~\bibnamefont{Cappelluti}},\ }%
  \bibfield{journal}{%
  \bibinfo {journal} {Phys. Rev. B}\ }%
  \textbf{\bibinfo {volume} {83}},\ \bibinfo {pages} {104516} (\bibinfo {year}
  {2011})%
  \bibAnnoteFile{NoStop}{BenfattoPRB11}%
\bibitem{FanfarilloPRB15}%
  \BibitemOpen
  \bibfield{author}{%
  \bibinfo {author} {\bibfnamefont{L.}~\bibnamefont{Fanfarillo}}, \bibinfo
  {author} {\bibfnamefont{A.}~\bibnamefont{Cortijo}},\ and\ \bibinfo {author}
  {\bibfnamefont{B.}~\bibnamefont{Valenzuela}},\ }%
  \bibfield{journal}{%
  \bibinfo {journal} {Phys. Rev. B}\ }%
  \textbf{\bibinfo {volume} {91}},\ \bibinfo {pages} {214515} (\bibinfo {year}
  {2015})%
  \bibAnnoteFile{NoStop}{FanfarilloPRB15}%
\bibitem{Watson2016}%
  \BibitemOpen
  \bibfield{author}{%
  \bibinfo {author} {\bibfnamefont{M.~D.}\ \bibnamefont{Watson}}
  \emph{et~al.},\ }%
  \bibinfo {journal} {arXiv:1603.04545}%
  \bibAnnoteFile{NoStop}{Watson2016}%
\bibitem{BorCondMat16}%
  \BibitemOpen
\bibfield{journal}{%
    }%
  \bibfield{author}{%
  \bibinfo {author} {\bibfnamefont{S.~V.}\ \bibnamefont{Borisenko}}
  \emph{et~al.},\ }%
  \bibfield{journal}{%
  \bibinfo {journal} {arxiV:1606.03022}}%
   (\bibinfo {year} {2016})%
  \bibAnnoteFile{NoStop}{BorCondMat16}%
\bibitem{Karlsson15}%
  \BibitemOpen
  \bibfield{author}{%
  \bibinfo {author} {\bibfnamefont{S.}~\bibnamefont{Karlsson}} \emph{et~al.},\
  }%
  \bibfield{journal}{%
  \bibinfo {journal} {Supercond. Sci. Technol.}\ }%
  \textbf{\bibinfo {volume} {28}},\ \bibinfo {pages} {105009} (\bibinfo {year}
  {2015})%
  \bibAnnoteFile{NoStop}{Karlsson15}%
\bibitem{AudouardEPL14}%
  \BibitemOpen
  \bibfield{author}{%
  \bibinfo {author} {\bibfnamefont{A.}~\bibnamefont{Audouard}}, \bibinfo
  {author} {\bibfnamefont{F.}~\bibnamefont{Duc}}, \bibinfo {author}
  {\bibfnamefont{L.}~\bibnamefont{Drigo}}, \bibinfo {author}
  {\bibfnamefont{P.}~\bibnamefont{Toulemonde}}, \bibinfo {author}
  {\bibfnamefont{S.}~\bibnamefont{Karlsson}}, \bibinfo {author}
  {\bibfnamefont{P.}~\bibnamefont{Strobel}},\ and\ \bibinfo {author}
  {\bibfnamefont{A.}~\bibnamefont{Sulpice}},\ }%
  \bibfield{journal}{%
  \bibinfo {journal} {EPL}\ }%
  \textbf{\bibinfo {volume} {109}},\ \bibinfo {pages} {27003} (\bibinfo {year}
  {2015})%
  \bibAnnoteFile{NoStop}{AudouardEPL14}%
\bibitem{EshrigPRB09}%
  \BibitemOpen
  \bibfield{author}{%
  \bibinfo {author} {\bibfnamefont{H.}~\bibnamefont{Eschrig}}\ and\ \bibinfo
  {author} {\bibfnamefont{K.}~\bibnamefont{Koepernik}},\ }%
  \bibfield{journal}{%
  \bibinfo {journal} {Phys. Rev. B}\ }%
  \textbf{\bibinfo {volume} {80}},\ \bibinfo {pages} {104503} (\bibinfo {year}
  {2009})%
  \bibAnnoteFile{NoStop}{EshrigPRB09}%
\bibitem{VafekPRB13}%
  \BibitemOpen
  \bibfield{author}{%
  \bibinfo {author} {\bibfnamefont{V.}~\bibnamefont{Cvetkovic}}\ and\ \bibinfo
  {author} {\bibfnamefont{O.}~\bibnamefont{Vafek}},\ }%
  \bibfield{journal}{%
  \bibinfo {journal} {Phys. Rev. B}\ }%
  \textbf{\bibinfo {volume} {88}},\ \bibinfo {pages} {134510} (\bibinfo {year}
  {2013})%
  \bibAnnoteFile{NoStop}{VafekPRB13}%
\bibitem{FernandesSO}%
  \BibitemOpen
  \bibfield{author}{%
  \bibinfo {author} {\bibfnamefont{R.~M.}\ \bibnamefont{Fernandes}}\ and\
  \bibinfo {author} {\bibfnamefont{O.}~\bibnamefont{Vafek}},\ }%
  \bibfield{journal}{%
  \bibinfo {journal} {Phys. Rev. B}\ }%
  \textbf{\bibinfo {volume} {90}},\ \bibinfo {pages} {214514} (\bibinfo {year}
  {2014})%
  \bibAnnoteFile{NoStop}{FernandesSO}%
\bibitem{Lee_pom_prb09}%
  \BibitemOpen
  \bibfield{author}{%
  \bibinfo {author} {\bibfnamefont{H.}~\bibnamefont{Zhai}}, \bibinfo {author}
  {\bibfnamefont{F.}~\bibnamefont{Wang}},\ and\ \bibinfo {author}
  {\bibfnamefont{D.-H.}\ \bibnamefont{Lee}},\ }%
  \bibfield{journal}{%
  \Doi{10.1103/PhysRevB.80.064517}{\bibinfo {journal} {Phys. Rev. B}}\ }%
  \textbf{\bibinfo {volume} {80}},\ \bibinfo {pages} {064517} (\bibinfo {month}
  {Aug}\ \bibinfo {year} {2009}),\
  \url{http://link.aps.org/doi/10.1103/PhysRevB.80.064517}%
  \bibAnnoteFile{NoStop}{Lee_pom_prb09}%
\bibitem{ChubukovCondMat16}%
  \BibitemOpen
  \bibfield{author}{%
  \bibinfo {author} {\bibfnamefont{A.~V.}\ \bibnamefont{Chubukov}}, \bibinfo
  {author} {\bibfnamefont{M.}~\bibnamefont{Khodas}},\ and\ \bibinfo {author}
  {\bibfnamefont{R.~M.}\ \bibnamefont{Fernadnez}},\ }%
  \bibinfo {journal} {cond-mat/1602.05503}%
  \bibAnnoteFile{NoStop}{ChubukovCondMat16}%
\bibitem{MaletzPRB14}%
  \BibitemOpen
\bibfield{journal}{%
    }%
  \bibfield{author}{%
  \bibinfo {author} {\bibfnamefont{J.}~\bibnamefont{Maletz}}, \bibinfo {author}
  {\bibfnamefont{V.~B.}\ \bibnamefont{Zabolotnyy}}, \bibinfo {author}
  {\bibfnamefont{D.~V.}\ \bibnamefont{Evtushinsky}}, \bibinfo {author}
  {\bibfnamefont{S.}~\bibnamefont{Thirupathaiah}}, \bibinfo {author}
  {\bibfnamefont{A.~U.~B.}\ \bibnamefont{Wolter}}, \bibinfo {author}
  {\bibfnamefont{L.}~\bibnamefont{Harnagea}}, \bibinfo {author}
  {\bibfnamefont{A.~N.}\ \bibnamefont{Yaresko}}, \bibinfo {author}
  {\bibfnamefont{A.~N.}\ \bibnamefont{Vasiliev}}, \bibinfo {author}
  {\bibfnamefont{D.~A.}\ \bibnamefont{Chareev}}, \bibinfo {author}
  {\bibfnamefont{A.~E.}\ \bibnamefont{Bohmer}}, \bibinfo {author}
  {\bibfnamefont{F.}~\bibnamefont{Hardy}}, \bibinfo {author}
  {\bibfnamefont{T.}~\bibnamefont{Wolf}}, \bibinfo {author}
  {\bibfnamefont{C.}~\bibnamefont{Meingast}}, \bibinfo {author}
  {\bibfnamefont{E.~D.~L.}\ \bibnamefont{Rienks}}, \bibinfo {author}
  {\bibfnamefont{B.}~\bibnamefont{Buchner}},\ and\ \bibinfo {author}
  {\bibfnamefont{S.~V.}\ \bibnamefont{Borisenko}},\ }%
  \bibfield{journal}{%
  \bibinfo {journal} {Phys. Rev. B}\ }%
  \textbf{\bibinfo {volume} {89}},\ \bibinfo {pages} {220506} (\bibinfo {year}
  {2014})%
  \bibAnnoteFile{NoStop}{MaletzPRB14}%
\bibitem{HaoPRX14}%
  \BibitemOpen
  \bibfield{author}{%
  \bibinfo {author} {\bibfnamefont{N.}~\bibnamefont{Hao}}\ and\ \bibinfo
  {author} {\bibfnamefont{J.}~\bibnamefont{Hu}},\ }%
  \bibfield{journal}{%
  \bibinfo {journal} {Phys. Rev. X}\ }%
  \textbf{\bibinfo {volume} {4}},\ \bibinfo {pages} {031053} (\bibinfo {year}
  {2014})%
  \bibAnnoteFile{NoStop}{HaoPRX14}%
\bibitem{BrouetLiFeAs}%
  \BibitemOpen
  \bibfield{author}{%
  \bibinfo {author} {\bibfnamefont{V.}~\bibnamefont{Brouet}}, \bibinfo {author}
  {\bibfnamefont{D.}~\bibnamefont{LeBoeuf}}, \bibinfo {author}
  {\bibfnamefont{P.-H.}\ \bibnamefont{Lin}}, \bibinfo {author}
  {\bibfnamefont{J.}~\bibnamefont{Mansart}}, \bibinfo {author}
  {\bibfnamefont{A.}~\bibnamefont{Taleb-Ibrahimi}}, \bibinfo {author}
  {\bibfnamefont{P.}~\bibnamefont{Le~Fevre}}, \bibinfo {author}
  {\bibfnamefont{F.}~\bibnamefont{Bertran}}, \bibinfo {author}
  {\bibfnamefont{A.}~\bibnamefont{Forget}},\ and\ \bibinfo {author}
  {\bibfnamefont{D.}~\bibnamefont{Colson}},\ }%
  \bibfield{journal}{%
  \bibinfo {journal} {Phys. Rev. B}\ }%
  \textbf{\bibinfo {volume} {93}},\ \bibinfo {pages} {085137} (\bibinfo {year}
  {2016})%
  \bibAnnoteFile{NoStop}{BrouetLiFeAs}%
\bibitem{LeePRL12}%
  \BibitemOpen
  \bibfield{author}{%
  \bibinfo {author} {\bibfnamefont{G.}~\bibnamefont{Lee}}, \bibinfo {author}
  {\bibfnamefont{H.~S.}\ \bibnamefont{Ji}}, \bibinfo {author}
  {\bibfnamefont{Y.}~\bibnamefont{Kim}}, \bibinfo {author}
  {\bibfnamefont{C.}~\bibnamefont{Kim}}, \bibinfo {author}
  {\bibfnamefont{K.}~\bibnamefont{Haule}}, \bibinfo {author}
  {\bibfnamefont{G.}~\bibnamefont{Kotliar}}, \bibinfo {author}
  {\bibfnamefont{B.}~\bibnamefont{Lee}}, \bibinfo {author}
  {\bibfnamefont{S.}~\bibnamefont{Khim}}, \bibinfo {author}
  {\bibfnamefont{K.~H.}\ \bibnamefont{Kim}}, \bibinfo {author}
  {\bibfnamefont{K.~S.}\ \bibnamefont{Kim}}, \bibinfo {author}
  {\bibfnamefont{K.-S.}\ \bibnamefont{Kim}},\ and\ \bibinfo {author}
  {\bibfnamefont{J.~H.}\ \bibnamefont{Shim}},\ }%
  \bibfield{journal}{%
  \bibinfo {journal} {Phys. Rev. Lett.}\ }%
  \textbf{\bibinfo {volume} {109}},\ \bibinfo {pages} {177001} (\bibinfo {year}
  {2012})%
  \bibAnnoteFile{NoStop}{LeePRL12}%
\bibitem{InosovNatPhys10}%
  \BibitemOpen
  \bibfield{author}{%
  \bibinfo {author} {\bibfnamefont{D.~S.}\ \bibnamefont{Inosov}}, \bibinfo
  {author} {\bibfnamefont{J.~T.}\ \bibnamefont{Park}}, \bibinfo {author}
  {\bibfnamefont{P.}~\bibnamefont{Bourges}}, \bibinfo {author}
  {\bibfnamefont{D.~L.}\ \bibnamefont{Sun}}, \bibinfo {author}
  {\bibfnamefont{Y.}~\bibnamefont{Sidis}}, \bibinfo {author}
  {\bibfnamefont{A.}~\bibnamefont{Schneidewind}}, \bibinfo {author}
  {\bibfnamefont{K.}~\bibnamefont{Hradil}}, \bibinfo {author}
  {\bibfnamefont{D.}~\bibnamefont{Haug}}, \bibinfo {author}
  {\bibfnamefont{C.~T.}\ \bibnamefont{Lin}}, \bibinfo {author}
  {\bibfnamefont{B.}~\bibnamefont{Keimer}},\ and\ \bibinfo {author}
  {\bibfnamefont{V.}~\bibnamefont{Hinkov}},\ }%
  \bibfield{journal}{%
  \bibinfo {journal} {Nat Phys}\ }%
  \textbf{\bibinfo {volume} {6}},\ \bibinfo {pages} {178} (\bibinfo {year}
  {2010})%
  \bibAnnoteFile{NoStop}{InosovNatPhys10}%
\bibitem{TanFengPRB16}%
  \BibitemOpen
  \bibfield{author}{%
  \bibinfo {author} {\bibfnamefont{S.~Y.}\ \bibnamefont{Tan}}, \bibinfo
  {author} {\bibfnamefont{Y.}~\bibnamefont{Fang}}, \bibinfo {author}
  {\bibfnamefont{D.~H.}\ \bibnamefont{Xie}}, \bibinfo {author}
  {\bibfnamefont{W.}~\bibnamefont{Feng}}, \bibinfo {author}
  {\bibfnamefont{C.~H.~P.}\ \bibnamefont{Wen}}, \bibinfo {author}
  {\bibfnamefont{Q.}~\bibnamefont{Song}}, \bibinfo {author}
  {\bibfnamefont{Q.~Y.}\ \bibnamefont{Chen}}, \bibinfo {author}
  {\bibfnamefont{W.}~\bibnamefont{Zhang}}, \bibinfo {author}
  {\bibfnamefont{Y.}~\bibnamefont{Zhang}}, \bibinfo {author}
  {\bibfnamefont{L.~Z.}\ \bibnamefont{Luo}}, \bibinfo {author}
  {\bibfnamefont{B.~P.}\ \bibnamefont{Xie}}, \bibinfo {author}
  {\bibfnamefont{X.~C.}\ \bibnamefont{Lai}},\ and\ \bibinfo {author}
  {\bibfnamefont{D.~L.}\ \bibnamefont{Feng}},\ }%
  \bibfield{journal}{%
  \bibinfo {journal} {Phys. Rev. B}\ }%
  \textbf{\bibinfo {volume} {93}},\ \bibinfo {pages} {104513} (\bibinfo {year}
  {2016})%
  \bibAnnoteFile{NoStop}{TanFengPRB16}%
\bibitem{LiShen15}%
  \BibitemOpen
  \bibfield{author}{%
  \bibinfo {author} {\bibfnamefont{W.}~\bibnamefont{Li}} \emph{et~al.},\ }%
  \bibinfo {journal} {arXiv:1509.01892}%
  \bibAnnoteFile{NoStop}{LiShen15}%
\bibitem{GallaisPRL13}%
  \BibitemOpen
\bibfield{journal}{%
    }%
  \bibfield{author}{%
  \bibinfo {author} {\bibfnamefont{Y.}~\bibnamefont{Gallais}}, \bibinfo
  {author} {\bibfnamefont{R.~M.}\ \bibnamefont{Fernandes}}, \bibinfo {author}
  {\bibfnamefont{I.}~\bibnamefont{Paul}}, \bibinfo {author}
  {\bibfnamefont{L.}~\bibnamefont{Chauvi\`ere}}, \bibinfo {author}
  {\bibfnamefont{Y.-X.}\ \bibnamefont{Yang}}, \bibinfo {author}
  {\bibfnamefont{M.-A.}\ \bibnamefont{M\'easson}}, \bibinfo {author}
  {\bibfnamefont{M.}~\bibnamefont{Cazayous}}, \bibinfo {author}
  {\bibfnamefont{A.}~\bibnamefont{Sacuto}}, \bibinfo {author}
  {\bibfnamefont{D.}~\bibnamefont{Colson}},\ and\ \bibinfo {author}
  {\bibfnamefont{A.}~\bibnamefont{Forget}},\ }%
  \bibfield{journal}{%
  \Doi{10.1103/PhysRevLett.111.267001}{\bibinfo {journal} {Phys. Rev. Lett.}}\
  }%
  \textbf{\bibinfo {volume} {111}},\ \bibinfo {pages} {267001} (\bibinfo
  {month} {Dec}\ \bibinfo {year} {2013}),\
  \url{http://link.aps.org/doi/10.1103/PhysRevLett.111.267001}%
  \bibAnnoteFile{NoStop}{GallaisPRL13}%
\bibitem{HacklNatPhys16}%
  \BibitemOpen
  \bibfield{author}{%
  \bibinfo {author} {\bibfnamefont{F.}~\bibnamefont{Kretzschmar}}, \bibinfo
  {author} {\bibfnamefont{T.}~\bibnamefont{Bohm}}, \bibinfo {author}
  {\bibfnamefont{U.}~\bibnamefont{Karahasanovic}}, \bibinfo {author}
  {\bibfnamefont{B.}~\bibnamefont{Muschler}}, \bibinfo {author}
  {\bibfnamefont{A.}~\bibnamefont{Baum}}, \bibinfo {author}
  {\bibfnamefont{D.}~\bibnamefont{Jost}}, \bibinfo {author}
  {\bibfnamefont{J.}~\bibnamefont{Schmalian}}, \bibinfo {author}
  {\bibfnamefont{S.}~\bibnamefont{Caprara}}, \bibinfo {author}
  {\bibfnamefont{M.}~\bibnamefont{Grilli}}, \bibinfo {author}
  {\bibfnamefont{C.}~\bibnamefont{Di~Castro}}, \bibinfo {author}
  {\bibfnamefont{J.~G.}\ \bibnamefont{Analytis}}, \bibinfo {author}
  {\bibfnamefont{J.~H.}\ \bibnamefont{Chu}}, \bibinfo {author}
  {\bibfnamefont{I.~R.}\ \bibnamefont{Fisher}},\ and\ \bibinfo {author}
  {\bibfnamefont{R.}~\bibnamefont{Hackl}},\ }%
  \bibfield{journal}{%
  \bibinfo {journal} {Nat Phys}\ }%
  \textbf{\bibinfo {volume} {12}},\ \bibinfo {pages} {560} (\bibinfo {month}
  {06}\ \bibinfo {year} {2016}),\ \url{http://dx.doi.org/10.1038/nphys3634}%
  \bibAnnoteFile{NoStop}{HacklNatPhys16}%
\bibitem{BlumbergPRB16}%
  \BibitemOpen
  \bibfield{author}{%
  \bibinfo {author} {\bibfnamefont{V.~K.}\ \bibnamefont{Thorsmolle}}, \bibinfo
  {author} {\bibfnamefont{M.}~\bibnamefont{Khodas}}, \bibinfo {author}
  {\bibfnamefont{Z.~P.}\ \bibnamefont{Yin}}, \bibinfo {author}
  {\bibfnamefont{C.}~\bibnamefont{Zhang}}, \bibinfo {author}
  {\bibfnamefont{S.~V.}\ \bibnamefont{Carr}}, \bibinfo {author}
  {\bibfnamefont{P.}~\bibnamefont{Dai}},\ and\ \bibinfo {author}
  {\bibfnamefont{G.}~\bibnamefont{Blumberg}},\ }%
  \bibfield{journal}{%
  \Doi{10.1103/PhysRevB.93.054515}{\bibinfo {journal} {Phys. Rev. B}}\ }%
  \textbf{\bibinfo {volume} {93}},\ \bibinfo {pages} {054515} (\bibinfo {month}
  {Feb}\ \bibinfo {year} {2016}),\
  \url{http://link.aps.org/doi/10.1103/PhysRevB.93.054515}%
  \bibAnnoteFile{NoStop}{BlumbergPRB16}%
\bibitem{KreiselCondMat15}%
  \BibitemOpen
  \bibfield{author}{%
  \bibinfo {author} {\bibfnamefont{A.}~\bibnamefont{Kreisel}}, \bibinfo
  {author} {\bibfnamefont{S.}~\bibnamefont{Mukherjee}}, \bibinfo {author}
  {\bibfnamefont{P.~J.}\ \bibnamefont{Hirschfeld}},\ and\ \bibinfo {author}
  {\bibfnamefont{B.~M.}\ \bibnamefont{Andersen}},\ }%
  \bibfield{journal}{%
  \bibinfo {journal} {Phys. Rev. B}\ }%
  \textbf{\bibinfo {volume} {92}},\ \bibinfo {pages} {224515} (\bibinfo {year}
  {2015})%
  \bibAnnoteFile{NoStop}{KreiselCondMat15}%
\bibitem{LiangCondMat15}%
  \BibitemOpen
  \bibfield{author}{%
  \bibinfo {author} {\bibfnamefont{Y.}~\bibnamefont{Liang}}, \bibinfo {author}
  {\bibfnamefont{X.}~\bibnamefont{Wu}},\ and\ \bibinfo {author}
  {\bibfnamefont{J.}~\bibnamefont{Hu}},\ }%
  \bibfield{journal}{%
  \bibinfo {journal} {Chinese Phys. Lett.}\ }%
  \textbf{\bibinfo {volume} {32}},\ \bibinfo {pages} {117402} (\bibinfo {year}
  {2015})%
  \bibAnnoteFile{NoStop}{LiangCondMat15}%
\bibitem{ZhaiPRB09}%
  \BibitemOpen
  \bibfield{author}{%
  \bibinfo {author} {\bibfnamefont{H.}~\bibnamefont{Zhai}}, \bibinfo {author}
  {\bibfnamefont{F.}~\bibnamefont{Wang}},\ and\ \bibinfo {author}
  {\bibfnamefont{D.-H.}\ \bibnamefont{Lee}},\ }%
  \bibfield{journal}{%
  \bibinfo {journal} {Phys. Rev. B}\ }%
  \textbf{\bibinfo {volume} {80}},\ \bibinfo {pages} {064517} (\bibinfo {year}
  {2009})%
  \bibAnnoteFile{NoStop}{ZhaiPRB09}%
\bibitem{KontaniCondMat15}%
  \BibitemOpen
  \bibfield{author}{%
  \bibinfo {author} {\bibfnamefont{S.}~\bibnamefont{Onari}}, \bibinfo {author}
  {\bibfnamefont{Y.}~\bibnamefont{Yamakawa}},\ and\ \bibinfo {author}
  {\bibfnamefont{H.}~\bibnamefont{Kontani}},\ }%
  \bibfield{journal}{%
  \bibinfo {journal} {Phys. Rev. Lett.}\ }%
  \textbf{\bibinfo {volume} {116}},\ \bibinfo {pages} {227001} (\bibinfo {year}
  {2016})%
  \bibAnnoteFile{NoStop}{KontaniCondMat15}%
\bibitem{Wien2k}%
  \BibitemOpen
  \bibfield{author}{%
  \bibinfo {author} {\bibfnamefont{P.}~\bibnamefont{Blaha}}, \bibinfo {author}
  {\bibfnamefont{K.}~\bibnamefont{Schwarz}}, \bibinfo {author}
  {\bibfnamefont{G.}~\bibnamefont{Madsen}}, \bibinfo {author}
  {\bibfnamefont{D.}~\bibnamefont{Kvasnicka}},\ and\ \bibinfo {author}
  {\bibfnamefont{J.}~\bibnamefont{Luitz}},\ }%
  \bibfield{journal}{%
  \bibinfo {journal} {WIEN2K: An Augmented Plane Wave + Local Orbitals Program
  for Calculating Crystal Properties (Karlheinz Schwarz, Techniche Universitat,
  Wien, Austria)}}%
   (\bibinfo {year} {1999})%
  \bibAnnoteFile{NoStop}{Wien2k}%
\bibitem{KhasanovNJP10}%
  \BibitemOpen
  \bibfield{author}{%
  \bibinfo {author} {\bibfnamefont{R.}~\bibnamefont{Khasanov}} \emph{et~al.},\
  }%
  \bibfield{journal}{%
  \bibinfo {journal} {New J. oh Physics}\ }%
  \textbf{\bibinfo {volume} {111}},\ \bibinfo {pages} {16309} (\bibinfo {year}
  {2014})%
  \bibAnnoteFile{NoStop}{KhasanovNJP10}%
\bibitem{BrouetPRB09}%
  \BibitemOpen
  \bibfield{author}{%
  \bibinfo {author} {\bibfnamefont{V.}~\bibnamefont{Brouet}}, \bibinfo {author}
  {\bibfnamefont{M.}~\bibnamefont{Marsi}}, \bibinfo {author}
  {\bibfnamefont{B.}~\bibnamefont{Mansart}}, \bibinfo {author}
  {\bibfnamefont{A.}~\bibnamefont{Nicolaou}}, \bibinfo {author}
  {\bibfnamefont{A.}~\bibnamefont{Taleb-Ibrahimi}}, \bibinfo {author}
  {\bibfnamefont{P.}~\bibnamefont{Le~F\`evre}}, \bibinfo {author}
  {\bibfnamefont{F.}~\bibnamefont{Bertran}}, \bibinfo {author}
  {\bibfnamefont{F.}~\bibnamefont{Rullier-Albenque}}, \bibinfo {author}
  {\bibfnamefont{A.}~\bibnamefont{Forget}},\ and\ \bibinfo {author}
  {\bibfnamefont{D.}~\bibnamefont{Colson}},\ }%
  \bibfield{journal}{%
  \bibinfo {journal} {Phys. Rev. B}\ }%
  \textbf{\bibinfo {volume} {80}},\ \bibinfo {pages} {165115} (\bibinfo {year}
  {2009})%
  \bibAnnoteFile{NoStop}{BrouetPRB09}%
\bibitem{BrouetPRB12}%
  \BibitemOpen
  \bibfield{author}{%
  \bibinfo {author} {\bibfnamefont{V.}~\bibnamefont{Brouet}}, \bibinfo {author}
  {\bibfnamefont{M.~F.}\ \bibnamefont{Jensen}}, \bibinfo {author}
  {\bibfnamefont{P.-H.}\ \bibnamefont{Lin}}, \bibinfo {author}
  {\bibfnamefont{A.}~\bibnamefont{Taleb-Ibrahimi}}, \bibinfo {author}
  {\bibfnamefont{P.}~\bibnamefont{Le~F\`evre}}, \bibinfo {author}
  {\bibfnamefont{F.}~\bibnamefont{Bertran}}, \bibinfo {author}
  {\bibfnamefont{C.-H.}\ \bibnamefont{Lin}}, \bibinfo {author}
  {\bibfnamefont{W.}~\bibnamefont{Ku}}, \bibinfo {author}
  {\bibfnamefont{A.}~\bibnamefont{Forget}},\ and\ \bibinfo {author}
  {\bibfnamefont{D.}~\bibnamefont{Colson}},\ }%
  \bibfield{journal}{%
  \bibinfo {journal} {Phys. Rev. B}\ }%
  \textbf{\bibinfo {volume} {86}},\ \bibinfo {pages} {075123} (\bibinfo {year}
  {2012})%
  \bibAnnoteFile{NoStop}{BrouetPRB12}%
\bibitem{MoreschiniPRL14}%
  \BibitemOpen
  \bibfield{author}{%
  \bibinfo {author} {\bibfnamefont{L.}~\bibnamefont{Moreschini}}, \bibinfo
  {author} {\bibfnamefont{P.-H.}\ \bibnamefont{Lin}}, \bibinfo {author}
  {\bibfnamefont{C.-H.}\ \bibnamefont{Lin}}, \bibinfo {author}
  {\bibfnamefont{W.}~\bibnamefont{Ku}}, \bibinfo {author}
  {\bibfnamefont{D.}~\bibnamefont{Innocenti}}, \bibinfo {author}
  {\bibfnamefont{Y.~J.}\ \bibnamefont{Chang}}, \bibinfo {author}
  {\bibfnamefont{A.}~\bibnamefont{Walter}}, \bibinfo {author}
  {\bibfnamefont{K.}~\bibnamefont{Kim}}, \bibinfo {author}
  {\bibfnamefont{V.}~\bibnamefont{Brouet}}, \bibinfo {author}
  {\bibfnamefont{K.-W.}\ \bibnamefont{Yeh}}, \bibinfo {author}
  {\bibfnamefont{M.-K.}\ \bibnamefont{Wu}}, \bibinfo {author}
  {\bibfnamefont{E.}~\bibnamefont{Rotenberg}}, \bibinfo {author}
  {\bibfnamefont{A.}~\bibnamefont{Bostwick}},\ and\ \bibinfo {author}
  {\bibfnamefont{M.}~\bibnamefont{Grioni}},\ }%
  \bibfield{journal}{%
  \bibinfo {journal} {Phys. Rev. Lett.}\ }%
  \textbf{\bibinfo {volume} {112}},\ \bibinfo {pages} {087602} (\bibinfo {year}
  {2014})%
  \bibAnnoteFile{NoStop}{MoreschiniPRL14}%
\bibitem{ZhangPRB11}%
  \BibitemOpen
  \bibfield{author}{%
  \bibinfo {author} {\bibfnamefont{Y.}~\bibnamefont{Zhang}}, \bibinfo {author}
  {\bibfnamefont{F.}~\bibnamefont{Chen}}, \bibinfo {author}
  {\bibfnamefont{C.}~\bibnamefont{He}}, \bibinfo {author}
  {\bibfnamefont{B.}~\bibnamefont{Zhou}}, \bibinfo {author}
  {\bibfnamefont{B.~P.}\ \bibnamefont{Xie}}, \bibinfo {author}
  {\bibfnamefont{C.}~\bibnamefont{Fang}}, \bibinfo {author}
  {\bibfnamefont{W.~F.}\ \bibnamefont{Tsai}}, \bibinfo {author}
  {\bibfnamefont{X.~H.}\ \bibnamefont{Chen}}, \bibinfo {author}
  {\bibfnamefont{H.}~\bibnamefont{Hayashi}}, \bibinfo {author}
  {\bibfnamefont{J.}~\bibnamefont{Jiang}}, \bibinfo {author}
  {\bibfnamefont{H.}~\bibnamefont{Iwasawa}}, \bibinfo {author}
  {\bibfnamefont{K.}~\bibnamefont{Shimada}}, \bibinfo {author}
  {\bibfnamefont{H.}~\bibnamefont{Namatame}}, \bibinfo {author}
  {\bibfnamefont{M.}~\bibnamefont{Taniguchi}}, \bibinfo {author}
  {\bibfnamefont{J.~P.}\ \bibnamefont{Hu}},\ and\ \bibinfo {author}
  {\bibfnamefont{D.~L.}\ \bibnamefont{Feng}},\ }%
  \bibfield{journal}{%
  \bibinfo {journal} {Phys. Rev. B}\ }%
  \textbf{\bibinfo {volume} {83}},\ \bibinfo {pages} {054510} (\bibinfo {year}
  {2011})%
  \bibAnnoteFile{NoStop}{ZhangPRB11}%
\bibitem{JensenPRB11}%
  \BibitemOpen
  \bibfield{author}{%
  \bibinfo {author} {\bibfnamefont{M.}~\bibnamefont{Fuglsang~Jensen}}, \bibinfo
  {author} {\bibfnamefont{V.}~\bibnamefont{Brouet}}, \bibinfo {author}
  {\bibfnamefont{E.}~\bibnamefont{Papalazarou}}, \bibinfo {author}
  {\bibfnamefont{A.}~\bibnamefont{Nicolaou}}, \bibinfo {author}
  {\bibfnamefont{A.}~\bibnamefont{Taleb-Ibrahimi}}, \bibinfo {author}
  {\bibfnamefont{P.}~\bibnamefont{Le~F\`evre}}, \bibinfo {author}
  {\bibfnamefont{F.}~\bibnamefont{Bertran}}, \bibinfo {author}
  {\bibfnamefont{A.}~\bibnamefont{Forget}},\ and\ \bibinfo {author}
  {\bibfnamefont{D.}~\bibnamefont{Colson}},\ }%
  \bibfield{journal}{%
  \bibinfo {journal} {Phys. Rev. B}\ }%
  \textbf{\bibinfo {volume} {84}},\ \bibinfo {pages} {014509} (\bibinfo {year}
  {2011})%
  \bibAnnoteFile{NoStop}{JensenPRB11}%
\bibitem{ZhangPRB12}%
  \BibitemOpen
  \bibfield{author}{%
  \bibinfo {author} {\bibfnamefont{Y.}~\bibnamefont{Zhang}}, \bibinfo {author}
  {\bibfnamefont{C.}~\bibnamefont{He}}, \bibinfo {author}
  {\bibfnamefont{Z.~R.}\ \bibnamefont{Ye}}, \bibinfo {author}
  {\bibfnamefont{J.}~\bibnamefont{Jiang}}, \bibinfo {author}
  {\bibfnamefont{F.}~\bibnamefont{Chen}}, \bibinfo {author}
  {\bibfnamefont{M.}~\bibnamefont{Xu}}, \bibinfo {author}
  {\bibfnamefont{Q.~Q.}\ \bibnamefont{Ge}}, \bibinfo {author}
  {\bibfnamefont{B.~P.}\ \bibnamefont{Xie}}, \bibinfo {author}
  {\bibfnamefont{J.}~\bibnamefont{Wei}}, \bibinfo {author}
  {\bibfnamefont{M.}~\bibnamefont{Aeschlimann}}, \bibinfo {author}
  {\bibfnamefont{X.~Y.}\ \bibnamefont{Cui}}, \bibinfo {author}
  {\bibfnamefont{M.}~\bibnamefont{Shi}}, \bibinfo {author}
  {\bibfnamefont{J.~P.}\ \bibnamefont{Hu}},\ and\ \bibinfo {author}
  {\bibfnamefont{D.~L.}\ \bibnamefont{Feng}},\ }%
  \bibfield{journal}{%
  \Doi{10.1103/PhysRevB.85.085121}{\bibinfo {journal} {Phys. Rev. B}}\ }%
  \textbf{\bibinfo {volume} {85}},\ \bibinfo {pages} {085121} (\bibinfo {month}
  {Feb}\ \bibinfo {year} {2012}),\
  \url{http://link.aps.org/doi/10.1103/PhysRevB.85.085121}%
  \bibAnnoteFile{NoStop}{ZhangPRB12}%
\bibitem{LeePRB09}%
  \BibitemOpen
  \bibfield{author}{%
  \bibinfo {author} {\bibfnamefont{Y.}~\bibnamefont{Ran}}, \bibinfo {author}
  {\bibfnamefont{F.}~\bibnamefont{Wang}}, \bibinfo {author}
  {\bibfnamefont{H.}~\bibnamefont{Zhai}}, \bibinfo {author}
  {\bibfnamefont{A.}~\bibnamefont{Vishwanath}},\ and\ \bibinfo {author}
  {\bibfnamefont{D.-H.}\ \bibnamefont{Lee}},\ }%
  \bibfield{journal}{%
  \Doi{10.1103/PhysRevB.79.014505}{\bibinfo {journal} {Phys. Rev. B}}\ }%
  \textbf{\bibinfo {volume} {79}},\ \bibinfo {pages} {014505} (\bibinfo {month}
  {Jan}\ \bibinfo {year} {2009}),\
  \url{http://link.aps.org/doi/10.1103/PhysRevB.79.014505}%
  \bibAnnoteFile{NoStop}{LeePRB09}%
\bibitem{Bascones16}%
  \BibitemOpen
  \bibfield{author}{%
  \bibinfo {author} {\bibfnamefont{E.}~\bibnamefont{Bascones}}, \bibinfo
  {author} {\bibfnamefont{B.}~\bibnamefont{Valenzuela}},\ and\ \bibinfo
  {author} {\bibfnamefont{M.~J.}\ \bibnamefont{Calderón}},\ }%
  \bibfield{journal}{%
  \Doi{http://dx.doi.org/10.1016/j.crhy.2015.05.004}{\bibinfo {journal}
  {Comptes Rendus Physique}}\ }%
  \textbf{\bibinfo {volume} {17}},\ \bibinfo {pages} {36 } (\bibinfo {year}
  {2016}),\ ISSN \bibinfo {issn} {1631-0705},\ \bibinfo {note} {iron-based
  superconductors / Supraconducteurs à base de fer},\
  \url{http://www.sciencedirect.com/science/article/pii/S1631070515000924}%
  \bibAnnoteFile{NoStop}{Bascones16}%
\end{thebibliography}%


\begin{widetext}
\Large
\textbf{Supplementary Information}
\vspace{0.4cm}

\noindent\normalsize
\noindent
\textbf{Comparison with LDA calculation}

\vspace{0.4cm}

In Fig. \ref{Fig_LDA}(a), we show the band structure we have calculated at $k_z$=0
(red) and $k_z$=1 (blue) using the Wien2k package \cite{Wien2k}. It is based on
the experimental structure of FeSe determined in Ref. \cite{KhasanovNJP10}, where the height of Se above the Fe plane is $z_{Se}$=1.46\AA (internal coordinate z=0.27). This
agrees very well with previously published band structure
\cite{EshrigPRB09,AichhornPRBFeSe,LiangCondMat15}, although we include here
spin-orbit coupling as well. The $xy$ hole band is expected to cross
the Fermi level in such a calculation and forms a rather large hole pocket. 

As often noted, the position of the $xy$ hole band is very sensitive to $z_{Se}$. Using $z_{Se}$=1.26\AA (z=0.23) in Fig. \ref{Fig_LDA}(b), it is now below $E_F$, as observed by ARPES. Although we see no reason why Se should move to this position, we use this calculation as a reference of the band structure without a $xy$ hole band. It is more likely that this is obtained in reality by renormalizing the hopping integral for $xy$ (in the TB model of Eshrig et al. \cite{EshrigPRB09}, this can for example be obtained by renormalizing the $t_{11}^{11}$ parameter by 2). The main features of the band structure are similar, although the size of the inner hole pocket increases in the second case and the size of the $k_z$ dependence on the $xz/yz$ electron pocket reduces (see also Fig. \ref{Fig_LDA}(c-d)). 

In both cases, we observe two doublets at M, formed by
$xz/yz$ bands and the two $xy$ bands, as expected by symmetry in the absence of nematic splitting \cite{FernandesSO}.

\vspace{0.4cm}

 \begin{figure*}[h]
 \centering
 \includegraphics[width=0.85\textwidth]{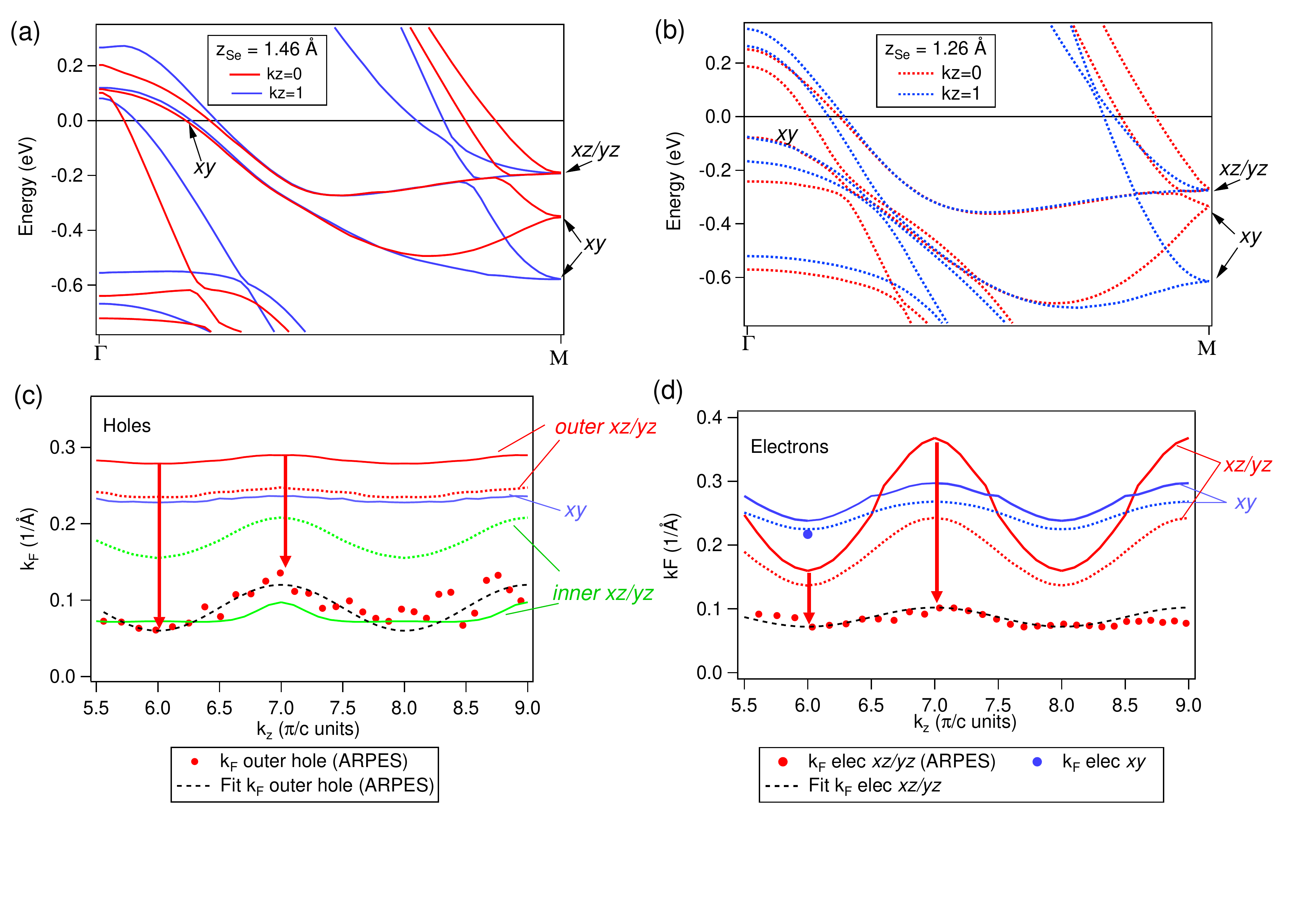}
 \caption{\textbf{Calculated band structure as a function of $k_z$.} (a)
 Calculated band structure using Wien2k at $k_z$=0 (red) and $k_z$=1 (blue) using the experimental structure, $z_{Se}$=1.46\AA. (b)
Same as (a) using an hypothetical structure with $z_{Se}$=1.26\AA to get the $xy$ hole band below $E_F$. (c) Calculated $k_F$ as a function of $k_z$ for the three hole bands
 in case (aà (solid lines) and (b) (dotted lines). Points display experimental $k_F$ for the outer $xz/yz$
 hole band. Big arrows indicate the shrinking at $k_z$=0 and 1. The inner hole
 band (green) is below E$_F$ at all $k_z$ experimentally. (c) Same for the
 electron pockets.} 
 \label{Fig_LDA}
 \end{figure*}

\noindent
\textit{k$_z$ dispersion and number of carriers - }
We measured the cuts of the paper\rq{}s Fig. 2 as a function of photon energy to check for three dimensional (3D)
effects. We observe a clear but rather small change of the Fermi wave vector
$k_F$ of the different pockets. From the periodicity of these changes, we can
deduce the value of $k_z$ at each photon energy, using the following formula
already heavily used in iron pnictides \cite{BrouetPRB09}. 
\begin{equation}
k_z=\sqrt{2m/\hbar^2*(h\nu-W+V_0)-k_{//}^2}
\label{Eq2}
\end{equation} 
We report $k_F$ for hole pockets in Fig. \ref{Fig_LDA}c and for electron
pockets in Fig. \ref{Fig_LDA}d as a function of $k_z$, using an inner
potential V$_0$=12eV. Comparison with the calculated sizes of the pockets (solid
lines for $z_{Se}$=1.46\AA, dotted lines for $z_{Se}$=1.26\AA) evidences a large FS shrinking in both cases. We could not observe the $xy$ electron
band on a sufficiently large photon energy window to determine its $k_z$
variation. We indicate the value $k_F$=0.217\AA$^{-1}$ determined at 40eV
($k_z \sim$0). 

For the outer hole band, $k_F$ evolves from 0.06 \AA$^{-1}$ at $k_z$=0 (38eV) to
0.12\AA$^{-1}$ at $k_z$=1 (54eV), in good agreement with results from Ref.
\cite{WatsonPRB15}. This corresponds to a number of carriers $n=\pi (k_Fa/\pi)^2/4$, assuming circular pockets, of 0.004 at $k_z$=0 and 0.016 at $k_z$=0, or 0.01 after integration over $k_z$. We find that the inner hole pocket is always below the Fermi
Surface, which is slightly different from Ref. \cite{WatsonPRB15}, where it
crosses $E_F$ in a small region near $k_z$=1. In any case, it contains a
negligible number of carriers. In the
calculations, the $xz/yz$ hole sheets contain nearly the same number of holes for the two cases considered in Fig. \ref{Fig_LDA}, namely 0.1, indicating a shrinking as large as 10.

For the $xz/yz$ electron band, $k_F$ changes from 0.072\AA$^{-1}$ at $k_z$=0
(42eV) to 0.102\AA$^{-1}$ at $k_z$=1 (58eV). At $k_z$=0, the reduction of $k_F$
is not as large as for the holes, but the very reduced $k_z$ dispersion finally
leads to a very large reduction of the number of electrons in the band. Considering only the $xz/yz$ parts of the electron sheets, we obtain 0.01 electrons experimentally, compared to 0.09 for the calculation with $z_{Se}$=1.46\AA~and 0.05 with $z_{Se}$=1.26\AA.~On the other hand, in both cases, there is almost no shrinking of the $xy$ electron band compared to both calculations.

\vspace{0.4cm}

\noindent
\textit{Renormalization - }
The discrepancy between the calculated and measured $k_z$ variation makes the
determination of the renormalization value somewhat ambiguous. Especially, the
choice of a renormalization value and of an applied shift are not independent
from each other. We choose to apply the same renormalization and shift at both
$k_z$ and the same renormalization for holes and electrons. In Fig. S2, we
compare the dispersions determined experimentally at $k_z$=0 (red) and 1 (blue)
by lorentzian fits of the Momentum Distribution Curves (MDC, circles) or maximum
of the Energy Distribution Curves (EDC, squares) with the LDA calculation at the
two $k_z$ (red and blue lines) for $z_{Se}$=1.46~\AA~(very similar values of renormalization and shift are obtained for $z_{Se}$=1.26~\AA, the main problem, in both cases, being disagreements on the $k_z$ dispersion). To obtain a reasonable agreement for hole bands,
we shift down the calculated bands by 200meV and renormalize them by 3. The
dotted line corresponds to the $xy$ hole band, after the same shift and
renormalization. Note that it is unlikely that $xy$ would be shifted and
renormalized in the same way as $xz/yz$, but the lack of experimental
information does not allow independent adjustment. The splitting between
$xz/yz$, due to SOC, is slightly larger in the calculation compared to
experiment. We attribute this to the influence of the $xy$ band, which increases
this splitting if it is located between $xz$ and $yz$. If $xy$ was not shifted
together with $xz/yz$, the SOC splitting would be reduced by 1.5 and would
indeed be very similar to the experimental one. 
 \begin{figure*}[tbp]
 \centering
 \includegraphics[width=1\textwidth]{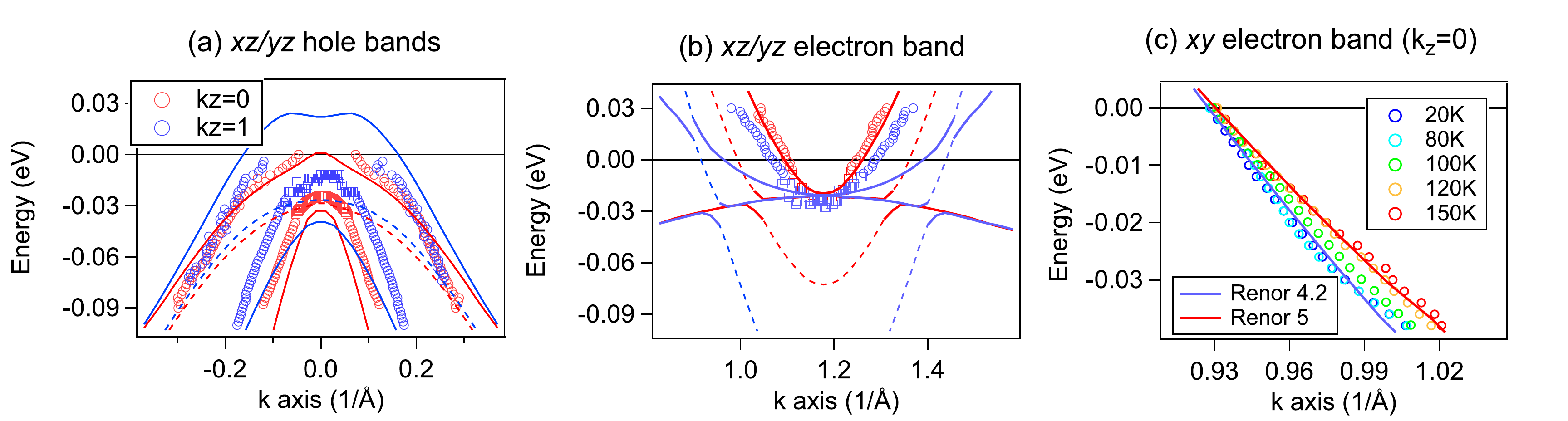}
 \caption{\textbf{Renormalized band structure.} (a) Points : experimental dispersion measured at $k_z$=0 (red) and $k_z$=1 (blue). Solid lines : calculated band structure for the two $k_z$, renormalized by 3 and shifted down by 0.2eV. The dotted line corresponds to the $xy$ hole band. (b) Same for electron pockets, with calculated bands renormalized by 3 and shifted up by 0.13eV. The dotted lines indicate the $xy$ electron band. (c) Initial slope of the $xy$ electron band at $k_z$=0 for different temperatures. Solid lines indicate calculation at $k_z$=0 renormalized by 4.2 (blue) or 5 (red).}
 \label{Fig_renor}
 \end{figure*}

For the $xz/yz$ electron bands, a renormalization value of 3 fixes an upward
shift of 130meV to fit the bottom of the band. This is a reliable point as it
does not change with $k_z$. As it was already clear from the very different
values of the shrinking in Fig. S1c, the fit cannot be simultaneously good for
$k_z$=0 and $k_z$=1.

For the $xy$ electron band, we show in Fig. \ref{Fig_renor}c the initial
slope of the dipsersion for different temperatures by open circles. The solid
line is the calculated $xy$ band renormalized by 4.2 (blue) and 5 (red). \\


%

\newpage
\noindent
\textbf{Detection of the $xy$ electron band}

In Fig. \ref{Sup_$xy$}a, we recall the selection rules for the different
bands. $\Gamma_2$ indicates the corner of the 1Fe BZ ($\pi$,$\pi$). We indicate by solid lines the
bands formed with the 2 Fe of the unit cell in-phase and by dotted lines those
formed by the 2 Fe out of phase (see Ref. \cite{BrouetPRB12}). As this changes
the parity of the bands with respect to the $xz$ mirror plane, this difference
is important to take into account, although it is not very often done. We
indicate the parities with respect to the plane containing the $\Gamma_1$M and z
axis, because selection rules will only be defined in the vicinity of this
scattering plane. In a calculation performed in a 1Fe BZ using glide mirror
symmetry, $xz$ and $yz$ would correspond to the solid lines and $xy$ to the
dotted lines, the difference being due to their different symmetry with respect
to $z$. Because of interference between the 2 Fe of the unit cell, bands shown
as dotted lines should generally give zero intensity in ARPES, unless they are
hybridized with bands of the opposite z parity. This indeed happens in many
cases and allows to detect some of these \lq\lq{}dotted\rq\rq{} bands quite
clearly (see Ref. \cite{BrouetPRB12} and \cite{MoreschiniPRL14}).

 \begin{figure}[tbp]
 \centering
 \includegraphics[width=0.9\textwidth]{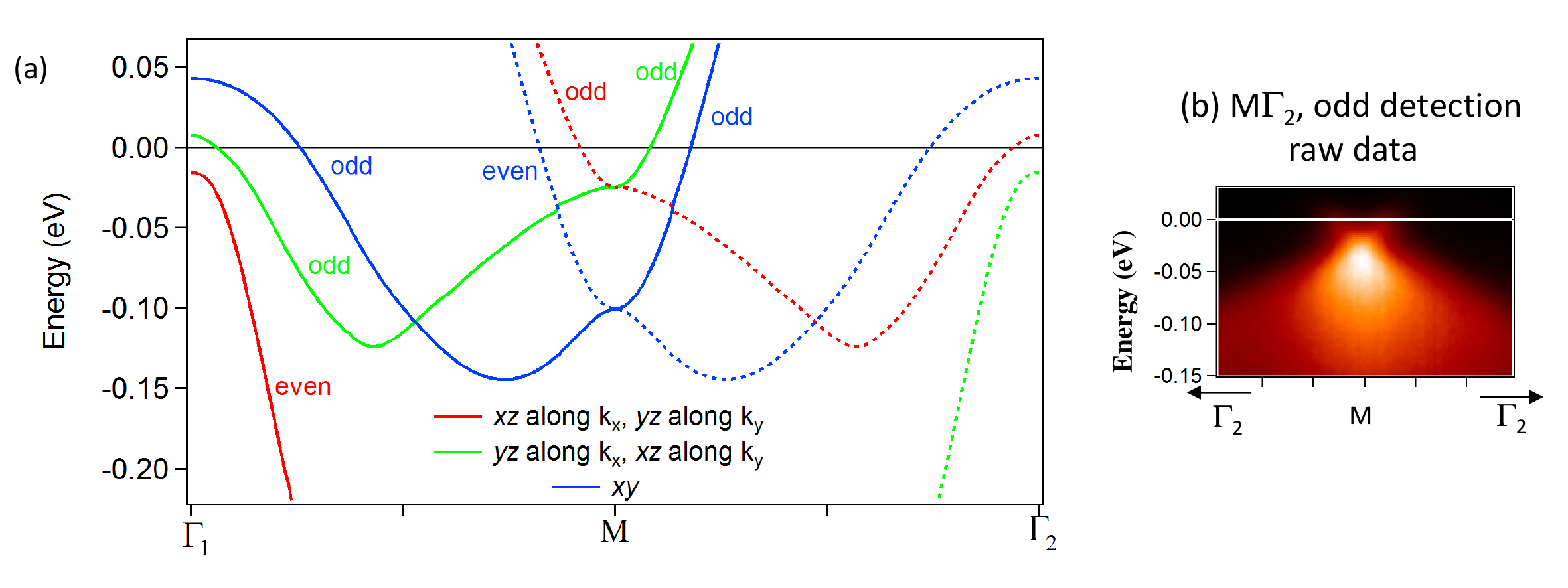}
 \caption{\textbf{Selection rules for the electron pockets.} (a) Sketch of the band structure of FeSe. The parity of the bands with respect to a plane containing $\Gamma_1$M and the surface normal is indicated. The dotted lines correspond to bands with the 2 Fe out of phase and generally have lower intensity in ARPES. (b) Raw image of the energy-momentum plot around M with odd detection at 135K. Normalizing the MDC to a constant area emphasizes the electron pocket [see Fig. 2(c)], while normalizing the EDC to a constant area emphasizes the saddle bands [see Fig. 4(a)].}
 \label{Sup_$xy$}
 \end{figure}

Most of the time, electron pockets are observed in the M$\Gamma_2$ direction
with odd configuration, which corresponds to cut (c) in our paper. This is
indeed the direction where the two electron bands should be strong. However,
most of the time the $xy$ electron pocket is not resolved due to a much lower
cross section for ARPES than $xz/yz$. Recently, Watson et al. claimed they could observe
the $xy$ band in this configuration at a particular photon energy
\cite{Watson2016}. For most photon energies, we only observed the shallow
$xz/yz$ band, as in Fig. 2c. Note that in the raw data, most of the weight is
concentrated at the bottom of the electron band, or equivalently, the top of the
saddle band in the perpendicular direction (see Fig. \ref{Sup_$xy$}b). To see
the electron band more clearly in Fig. 2c, we normalized all MDC spectra to the
same area. Alternatively, if all EDC spectra are normalized to the same area, as
we do in Fig. 4(b), we increase the weight of the saddle band. In
the perpendicular direction $\Gamma_1$M, keeping odd detection, the saddle band
is much stronger and the electron band is much weaker, as expected from
interference effects and shown in cut (a) of the paper.

 \begin{figure}[tbp]
 \centering
 \includegraphics[width=0.9\textwidth]{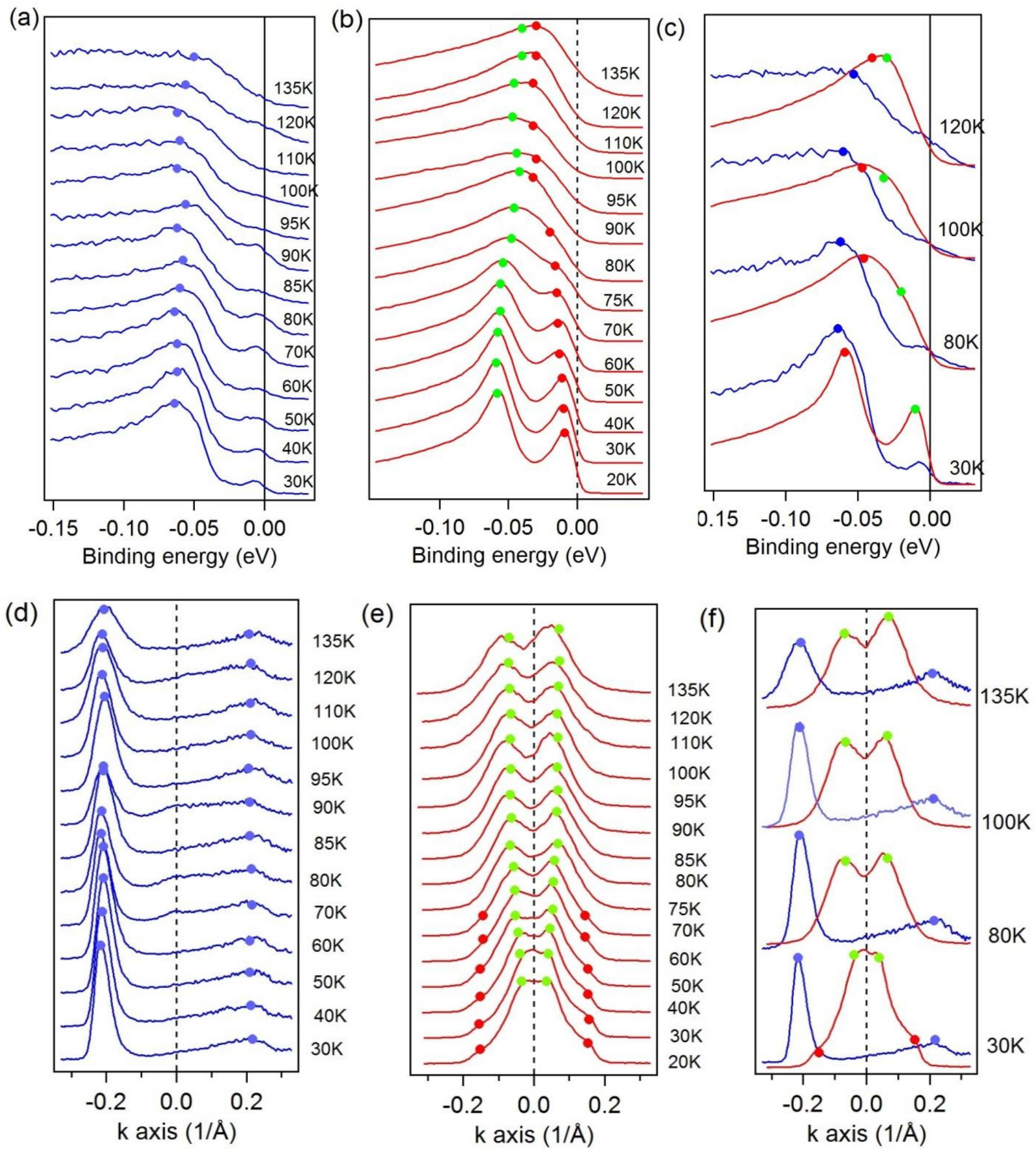}
 \caption{\textbf{Temperature evolution of the EDCs and MDCs} (a-c) EDC taken at M for the $xy$ (blue) and $xz/yz$ (red) electron bands at the indicated temperatures. Some spectra are superposed for direct comparisons in (c). Points are reproduced in Fig. 4 of the paper. (d-f) Same for MDCs at $E_F$. }
 \label{Sup_Stacks}
 \end{figure}

Along $\Gamma_1$M in even configuration (cut (b) in the paper), there is no
strong band expected around M. This allows to detect quite clearly the isolated
$xy$ band, despite the fact it is built from out of phase Fe, thanks to a
significant hybridization with $xz$ at some $k_z$ (see Fig. 6 in ref.
\cite{BrouetPRB12}). This configuration allows to isolate $xy$ in
Ba(Fe,Co)$_2$As$_2$ \cite{BrouetPRB12,ZhangPRB11}, BaFe$_2$As$_2$
\cite{JensenPRB11}, LiFeAs \cite{BrouetLiFeAs} or NaFeAs \cite{ZhangPRB12}. This also works for FeSe, as
shown in Fig. 4(b), but this configuration was never explicitly
used so far to study this band to our knowledge. The asymmetric intensity of
this band and the weak intensity at the bottom is a characteristic common to all
iron pnictides we have studied.

Comparing the two situations clarifies the evolution of the bands with
temperature. We report in Fig. 4(a) the model of the $xy$ band extracted from Fig. 4(b). The
bottom of the two bands are closer than in the calculation, due to the stronger
renormalization of $xy$, but they still can be clearly separated. Direct comparison of the EDC at M is given in
Fig. \ref{Sup_Stacks}. Although there is a small contribution of the red spectra in the blue one, the difference of position and width of the bottom band is clear. The two $xy$ bands could split if there is a significant
anisotropy of $xy$ hopping along $k_x$ and $k_y$ \cite{FernandesSO}, but this
would not be expected at high temperatures. Similarly, the strong temperature
evolution of the lower saddle band in Fig. 4(a) is totally uncorrelated with a similar
motion of the bottom of the $xy$ band. This supports the idea of a $xz/yz$
splitting with temperature. The weak electron band observed at 30K in (c) with a
bottom at -50meV is close to $xy$, but distinct. It is assigned to $xz$ and
further confirms the $xz/yz$ splitting. The
evolution of the bottom of the bands and their $k_F$ are indicated on the stacks
of Fig. \ref{Sup_Stacks} and reported in the paper's Fig. 3(e-f). \\

\noindent
\textbf{Self-energy corrections in the orbital-selective spin-fluctuations scenario}

Let us start from the model describing the bare electronic structure at each
pocket $l$, $H_0^l=\sum_{\bk,\s} \Psi^{l,\dagger}_{\bk\s} \hat H_0^l
\Psi^l_{\bk\s}$ where $\hat H_0^l$ ($l=\G,M_X,M_Y$) is a $2\times2$ matrix and
the spinors are defined as $\Psi^\G_{\bk\s}=(c^{yz}_{\bk\s},c^{xz}_{\bk\s})$ and
$\Psi^{X/Y}_{\bk\s}=(c^{yz/xz}_{\bk\s},c^{xy}_{\bk\s})$.  The matrix $\hat H^l$
has the general structure of Eq. (1) of the main text: %
\be
\lb{h0gen}
\hat H_0^{l} = h_0^l\t_0+\vec{h}^l\cdot\vec{\t}^l = \begin{pmatrix}
h_0^l(\bk)+h_3^l(\bk) & h_1^l(\bk)-ih_2^l(\bk) \\
h_1^l(\bk)+ih_2^l(\bk)& h_0(\bk)-h_3^l(\bk)\\
\end{pmatrix}
\ee
where $\vec{\t}$ are Pauli matrices representing the orbital isospin. The
eigenevalues $E^{l\pm}_\bk$ of the matrix \pref{h0gen} define the band
dispersion at each pocket, 
\be
\lb{epm}
E^{l\pm}_\bk=h_0^l\pm|\vec h^l|=h_0^l\pm\sqrt{(h^l_1)^2+(h^l_2)^2+(h^l_3)^2}.
\ee
The corresponding eigenvectors $\Psi^{l,+}=(u^l_\bk,v^l_\bk)$, $\Psi^{l,-}=(-v^l_\bk,u^l_\bk)$, where
\bea
\lb{u0}
(u^l_\bk)^2&=&\frac{1}{2}\left( 1+\frac{h^l_{3\bk}}{|\vec h^l_\bk|}\right),\\
\lb{v0}
(v^l_\bk)^2&=&\frac{1}{2}\left( 1-\frac{h^l_{3\bk}}{|\vec h^l_\bk|}\right).
\eea
define the rotation $\hat U^l=(\Psi^{l,+}, \Psi^{l,-})$ to the band basis, that
is needed to identify the orbital character of the bands. Indeed, the bare
Greens' function of the Hamiltonian \pref{h0gen} is given by
$G^l=U^l(i\o_n-\L^l)^{-1}U^{l\dagger}$, so that 
\be
G^l=g_+^l(i\o_n)
\begin{pmatrix}
 (u^l)^2& u^lv^l\\
u^lv^l & (v^l)^2\\
\end{pmatrix}+
\lb{grot}
g_-^l(i\o_n)
\begin{pmatrix}
(v^l)^2 & -u^l v^l\\
-u^l v^l & (u^l)^2\\
\end{pmatrix},
\ee
where we introduced the pocket's Green's function $g_\pm^l(i\o_n)= (i\o_n-E^{l \pm}_\bk)^{-1}$.
%
%

Since the bare model \pref{h0gen} is intended to include already the high-energy
renormalization of the bands observed in ARPES, we define the $h^l_i(\bk)$
functions following the general expressions of Ref.\ \cite{VafekPRB13}, with 
parameter values appropriate to describe the ARPES dispersions at 150 K, apart
from the band shift. We then have at the $\G$ pocket: 
\bea
h^\G_{0}(\bk)&=&\e_\G-a_\G \bk^2,\nonumber\\
h^\G_{1}(\bk)&=&-2b_\G k_xk_y,\nonumber\\
h^\G_{2}(\bk)&=&0,\nonumber\\
h^\G_{3}(\bk)&=&b_\G(k_x^2-k_y^2).
\eea
At the $\G$ point we should also add the spin-orbit coupling (SOC) $\lambda$
that acts with a term $\pm(\l/2)\tau_2$ on the $xz$ and $yz$ subspace for each
spin projection. More specifically, if one defines the enlarged spinor
$\Psi^\G_{\bk}=(c^{yz}_{\bk\up},c^{xz}_{\bk\up},c^{yz}_{\bk\down},c^{xz}_{
\bk\down})$ the SOC gives an additional term $(\l/2) \tau_2 \otimes \sigma_z$,
where $\sigma_z$ is the Pauli matrix acting on the spin sector\cite{VafekPRB13}.
One can easily see that the corresponding eigenvalues and eigenvectors preserve
the same structure given above, provided that one identifies at the $\G$ point
$h_2^\G$ with $\l/2$. As a consequence the two eigenvalues \pref{epm} read: 
\bea
\lb{epmbare}
E^{\G_\pm}_\bk&=&\e_\G-a_\G \bk^2\pm\sqrt{b_\G^2\bk^4+\l^2/4},
\eea
giving two parabolic bands splitted by the SOC at $\G$, $E^{\G_-}$ being the
inner hole pocket and $E^{\G_+}$ the outer one. At energies larger than SOC one
can also  see that the functions \pref{u0}-\pref{v0} become simply
$(u^\G)^2\simeq\cos^2 \theta_\bk$ and $(v^\G)^2\simeq\sin^2 \theta_\bk$ with
$\tan\theta_\bk=k_y/k_x$, giving the typical orbital content of the FS cut shown
in Fig. 1b,c of the manuscript. At the $X$ pocket the low-energy model reads:
\bea
h_0^X(\bk)&=&\frac{h_{yz}-h_{xy}}{2}\\
h_3^X(\bk)&=&\frac{h_{yz}-h_{xy}}{2}-b(k_x^2-k_y^2)\\
h_1^X(\bk)&=&0\\
h_2^X(\bk)&=&v k_y
\eea
%
with $h_{yz}=-\e_{yz}+a_{yz}k^2$ and $h_{xy}=-\e_{xy}+a_{xy}k^2$, so that
$E^{X+}$ is an electron-like band crossing the Fermi level, and $E^{X-}$ a
hole-like band remaining always below the Fermi level. The dispersion at the $Y$
pocket is obtained by exchanging $k_x\ra k_y$ in the above expressions, and with
the $yz$ label intended as a $xz$ label, due to the different orbital content at
$Y$. By measuring the momenta in units of the inverse lattice spacing one
obtains the parameter values listed in table \ref{table-par}. The resulting band
dispersions are shown by dotted lines in Fig. 1 of the manuscript. \\
\begin{table}
\caption{Low-energy model parameters in meV.}
\begin{tabular}{|c|c|}
\hline
\phantom{XX} $\G$ pocket\phantom{XX}& \phantom{XX}$X$ pocket \phantom{XX}\\
\hline
$\e_\G$=47 &$\e_{xy}=\e_{yz}$=57 \\
$a_\G$=127 &$a_{xy}$=225, \,$a_{yz}$=525 \\
$b_\G$=61 &$b_X$=725\\
 &$v$=240\\
\hline
\end{tabular}
\label{table-par}
\end{table}

To compute the role of spin fluctuations we start from the following microscopic
interacting Hamiltonian\cite{FanfarilloPRB15} 
\be
\lb{hint}
H_{int}=-1/2\sum_{\bq \\'}U_{\eta\eta'} \bS^\eta_{\bq } \cdot \bS^{\eta'}_{-\bq},
\ee
with $U_{\eta\eta'}=8/3U\d_{\eta\eta'}+4J_H(1-\delta_{\eta\eta'})$,
$\bS^\eta_{\bq }=\sum_{\bk ss'}(c^{\eta \dagger}_{\bk s} \s_{ss'}c^\eta_{\bk+\bq
s'})$ and $\s_{ss'}$ the Pauli matrices for the spin.  Here $\eta,\eta'=yz,xz,xy$
denote the orbital index. Notice that in the spin operator we take into account
only the intraorbital contribution, that is expected to be the dominant one\cite{LeePRB09,Bascones16}, but
the Hamiltonian \pref{hint} includes also inter-orbital spin-spin interaction
terms. 
The role of the orbital degrees of freedom for the description of the spin
fluctuations (SF) has been already discussed in Ref.\ \cite{FanfarilloPRB15}.
Here it has been shown that the SF peaked around the two vectors $\bQ_X$ or
$\bQ_Y$ involve only the $yz$ or $xz$ orbitals, respectively. {This result
basically follows from the fact that in the 1Fe BZ, only $yz$ is present in the
electron pocket at $\bQ_X$, and only $xz$ is present at $\bQ_Y$.} For example,
SF around $\bQ_X$ involve the two hole-like $\G$ pockets ($g^\G_\pm$) and the
electron-like $X$ pocket ($g^X_+$), but their contribution is weighted with the
$yz$ orbital content, i.e. only $\langle \bS^{yz}\cdot \bS^{yz}\rangle$ can be
different from zero. In a short notation, one recognizes that 
\bea
\lb{corrx}
\langle \bS\cdot \bS\rangle (\bQ_X) &\Ra& \langle \bS^{yz}_{\bQ_X}\cdot \bS^{yz}_{\bQ_X}\rangle \\
\lb{corry}
 \langle \bS\cdot \bS\rangle (\bQ_Y) &\Ra& \langle \bS^{xz}_{\bQ_Y}\cdot \bS_{\bQ_Y}^{xz}\rangle   
\eea
where e.g. the $\bS^{yz}_{\bQ_X}$ operator creates a particle-hole pair made by
$yz$ states at $\G$ and $X$,  and analogously for the $\bQ_Y$ operator. More
specifically, from Eq.\ \pref{grot} one sees that the $yz$ orbital content is
given by $(u^\G)^2$ for the outer pocket, by $(v^\G)^2$ for the inner $\G$
pocket and by $(u^X)^2$ for the electron pocket. Thus the SF propagator around
$\bQ_X$ is built with the following bare susceptibilities: 
\bea
\chi^{0}_{\Gamma_+ X}(q)&=& \frac{1}{4} \sum_{k} \, (u^{\G}_{\bk})^2 (u^X_{\bk+\bq})^2 \ g^{\Gamma}_+(k) g^{X}_+(k+q),\nn\\
\chi^{0}_{\Gamma_- X}(q)&=&\frac{1}{4}  \sum_{k} \, (v^{\G}_{\bk})^2 (u^X_{\bk+\bq})^2\ \ g^{\Gamma}_-(k) g^{X}_+(k+q),
\label{barechi}
\eea
where $k,q$ denote both the Matsubara frequency and the momentum. The bare SF
propagator around $\bQ_Y$ has an analogous definition, provided that $u^\G\ra
v^\G$, $u^X\ra u^Y$  and $g^X_+\ra g^Y_+$. By making the usual RPA resummation
one then finds\cite{FanfarilloPRB15}: 
\bea
\lb{chi+x}
\chi_{\Gamma_+ X}(q)&=& \frac{\chi^{0}_{\Gamma_+ X}(q)}{3/8U+\chi^{0}_{\Gamma_+ X}(q)}\\
\lb{chi-x}
\chi_{\Gamma_- X}(q)&=& \frac{\chi^{0}_{\Gamma_- X}(q)}{3/8U+\chi^{0}_{\Gamma_- X}(q)}
\eea
Apart from the presence of the modulating orbital factors in the bare bubbles (Eq.s \pref{barechi}),
one recognizes in Eq.s\ \pref{chi+x}-\pref{chi-x} the usual expression  for the
SF susceptibility of two quasi-nested bands, that diverges at zero momenta and
frequency when the band dispersions are perfectly nested, i.e.
$E^\G(\bk)=-E^X(\bk)$. 

A second consequence of the orbital-selectivity of the spin fluctuations is that
the self-energy matrix computed with the SF \pref{corrx}-\pref{corry} acquires
the diagonal form of Eq. (2) of the manuscript, and the orbital self-energies
are given explicitly by: 
\bea
\Sigma^{\G }_{yz}&=&g^2\sum_q G^{X}_{yzyz}(k-q) \langle \bS_{\bQ_X+\bq }^{yz} \cdot \bS_{\bQ_X-\bq }^{yz} \rangle
=g^2\sum_q (u^X_{\bk-\bq})^2 g^X_+(k-q)(\chi_{\G_+ X}(q)+\chi_{\G_- X}(q)).\nn \\
\Sigma^{\G }_{xz}&=&g^2\sum_q G^{Y}_{xzxz}(k-q) \langle \bS_{\bQ_Y+\bq }^{yz} \cdot \bS_{\bQ_Y-\bq }^{yz} \rangle
=g^2\sum_q (u^Y_{\bk-\bq})^2 g^Y_+(k-q)(\chi_{\G_+ Y}(q)+\chi_{\G_- Y}(q)). \nn \\
\lb{s11}
\eea
where the $\chi_{\G_\pm X/Y}$ have been defined in Eq.s
\pref{chi+x}-\pref{chi-x} above, and we put $g^2=(8U/3)^2/2$. 

When we compute instead the self-energy for the $X/Y$ pockets only the
$\Sigma_{yz/xz}$ components are corrected. With the same reasoning as before one
easily sees that 
\be
\Sigma^{X}_{yz}=g^2\sum_q G^{\G}_{yzyz}(k-q) \langle \bS_{\bQ_x+\bq }^{yz} \cdot \bS_{\bQ_x-\bq }^{yz} \rangle=
g^2\sum_q (u^\G_{\bk-\bq})^2 g^\G_+(k-q)\chi_{\G_+ X}(q)+ g^2\sum_q (v^\G_{\bk-\bq})^2 g^\G_-(k-q)\chi_{\G_- X}(q),
\lb{s22}
\ee
while at $Y$ only $\Sigma^Y_{xz}\neq 0$ with an expression analogous to Eq.\
\pref{s22} provided that $u^\G\leftrightarrow v^\G$ and $\chi_{\G_\pm X}\ra
\chi_{\G_\pm Y}$. To simplify the numerical computation of the self-energy
corrections and to present results valid for general band dispersions we discard
in the following the full momentum dependence of the SF propagators $\chi_{\G\pm
X/Y}$, and we retain only their frequency dependence. Indeed, as 
discussed previously\cite{OrtenziPRL09,BenfattoPRB11}, this is enough to capture 
the basic ingredients of the FS shrinking. Moreover we discard
possible differences between $\chi_{\G_+X/Y}$ and $\chi_{\G_-X/Y}$ due to the
different nesting conditions between the outer/inner hole and electron pockets,
but we retain the effect due to the mismatch of the angular factors at different
pockets in Eq.\ \pref{chi+x}-\pref{chi-x}. 
This can be roughly estimated from
the angular overlap of the orbital functions, approximated as $(u^\G)\sim \cos
\theta_\bk^2$,  $(v^\G)\sim \sin \theta_\bk^2$ and $(u^X)\sim \sin
\theta_\bk^2$. Then we can finally put 
\bea
\lb{self}
\Sigma^\G_{yz}(i\omega_n)&=&
-V^X_{he}T\sum_{m}
D_X(\omega_n-\omega_m)
g^X_+(i\omega_m),\nonumber \\
\Sigma^\G_{xz}(i\omega_n)&=&
-V^Y_{he}T\sum_{m}
D_Y(\omega_n-\omega_m)
g^Y_+(i\omega_m),\nonumber\\
\eea
and at the electron pockets
\bea
\Sigma^{X/Y}_{yz/xz}(i\omega_n)&=&
-V^{X/Y}_{eh}T\sum_{m}
D_{X/Y}(\omega_n-\omega_m) \left(\frac{1}{2}
g^\G_+(i\omega_m)+\frac{3}{8}g^\G_-(i\omega_m)\right)
\eea
where $g^l_{\pm}(i\o_n)=\sum_\bk g_\pm^l(\bk,i\omega_n)$ is the local Green's
function for the $E^{l,\pm}$ band  and $D_{X/Y}(\omega_l)= \int d\Omega 2\Omega
B_{XY}(\Omega)/(\Omega^2+\omega_l^2)$ is the bosonic propagator, with the
$B_{X/Y}$ spectral functions defined in Eq.\  (5) of the main text, i.e. 
\be
\lb{bapp}
B_{X,Y}(\o)=\frac{1}{\pi}\frac{\o \, \o_{0}}{(\o^{X/Y}_{sf}(T))^2+\O^2}.
\ee
As already discussed in the main text, the real parts of 
$\Sigma^{\G}_{yz/xz}$ are negative, as due to the electron-like particle-hole 
asymmetry of the electron pockets appearing on the r.h.s. of Eq.\ \pref{self}, 
and conversely those of $\Sigma^{X/Y}_{yz/xz}$ are positive.

The typical energy scale of the spin modes is set by $\o_0=20$ 
meV in Eq.\ \pref{bapp}.The $\omega^{X/Y}_{sf}(T)$ are the temperature-dependent 
masses for the SF along $k_x$ or $k_y$ shown in Fig. 5a of the manuscript. Above 
$T_c$ we use
$$\o^{X/Y}_{sf} (T)= \o_0 \,  h(T), \quad T>T_s,$$ 
as observed in iron-based systems \cite{InosovNatPhys10,Rahn15}. 
Below $T_S$ the anisotropy of SF is described by a splitting of the SF masses 
$\o_{sf}^X$ and $\o_{sf}^Y$, making SF stronger in the $k_x$ direction were the 
spin modes soften: 
$$\o^{X/Y}_{sf} (T) = \o_0 \, h(T_s)\big( 1 \mp \alpha \, g(T/T_s)\big),  \quad T<T_s.$$
Here $h(T)= (T + T_{\theta})/T_{\theta}$, $g(x)= (1 - x^4/3)\sqrt{1 - x^4}$, 
with $\alpha = 0.25$. At 150 K the couplings are chosen to reproduce the 
experimental band shifts, giving $V^{X/Y}_{eh}=26$ meV and $V^{X/Y}_{he}=38$ 
meV. Below $T_S$ the SF become stronger at $\bQ_X$. This reflects both in a 
softening of the $\omega_{sf}^X$ and in an enhancement of $V_{eh/he}^X$ with 
respect to the analogous quantities along $k_y$. For the coupling we assume the 
analogous $T$ evolution of the spin masses i.e. we put $V^{X/Y}_{eh/he} (T) \sim 
h^{-1}(T)$ above $T_s$ and $V^{X/Y}_{eh/he} (T) \sim \big(h(T_s)(1 \mp 
\beta^{X/Y}_{eh/he} \, g(T/T_s))\big)^{-1}$ below $T_s$ with $\beta^{X}_{eh} = 
0.49$, $\beta^{X}_{he} = 0.40$, $\beta^{Y}_{eh} = 0.25$, $\beta^{Y}_{he} = 
3.17$.

Observe that the self-energy matrix at the $\G$ point can be written in general as
\be
\lb{smatrixG}
\hat \Sigma^\G=\begin{pmatrix}
\Sigma^\G_{yz} & 0\\
0 &\Sigma^\G_{xz} \\
\end{pmatrix}=\Sigma^\G_0\t_0+\Delta\Sigma^\G\t_3,
\ee
with $\Sigma_0^\G=\frac{\Sigma^\G_{yz}+\Sigma^\G_{xz}}{2}$ and
$\Delta\Sigma^\G=\frac{\Sigma^\G_{yz}-\Sigma^\G_{xz}}{2}$. As a consequence, one
can directly see that the renormalized bands are defined by the maxima of the
spectral functions 
\bea
\lb{eigenv}
A^\G_\pm&=&-\frac{1}{\pi} Im \frac{1}{\omega-E^{R,\G}_\pm(\o)}\\
E^{R,\G}_\pm(\o)&=&h_0^\G+\Sigma_0^\G(\o)\pm\sqrt{(h^\G_3+\Delta\Sigma(\o))^2+(h^\G_1)^2+\l^2/4}\nn
\eea
In the paramagnetic phase $\Sigma^\G_{yz}=\Sigma^\G_{xz}$ and the $\t_3$ term is
absent, so that the two hole pockets are both shifted by the real part of the
$\Sigma_0^\G$ function, in full analogy with the result obtained before in the
band language\cite{OrtenziPRL09}. In contrast in the nematic phase where
$\Sigma^\Gamma_{yz}\neq \Sigma^\Gamma_{xz}$ and the $\tau_3$ correction is
present the self-energy corrections act as a crystal field that splits the bands
and changes the orbital character of the bands. The SO coupling contribute to
change the orbital character further. Indeed also the orbital factors
\pref{u0}-\pref{v0} are corrected as 
\bea
(u^\G_{R\bk})^2&=&\frac{1}{2}\left( 1+\frac{h_{3\bk}^\G+{Re}\Delta\Sigma^\G_\bk}{h^\G_{R\bk}}\right),\\
(v^\G_{R\bk})^2&=&\frac{1}{2}\left( 1-\frac{h_{3\bk}^\G+{Re} \Delta\Sigma^\G_\bk}{h^\G_{R\bk}}\right).
\eea
where $h^\G_{R\bk}$ denotes the square-root in Eq.\ \pref{eigenv}. 
 
Analogously, at the $X/Y$ pockets one sees that the new bands are defined by the spectral functions
\bea
A^X_\pm&=&-\frac{1}{\pi} Im \frac{1}{\omega-E^{R,X}_\pm(\o)}\\
E^{R,X}_\pm(\o)&=&h_0^X+\frac{\Sigma^X(\o)}{2}\pm\sqrt{(h^X_3+\Sigma^X(\o)/2)^2+(h^X_2)^2}.\nn\\
\eea
\end{widetext}
\end{document}